\documentclass[%
 aip,
 amsmath,amssymb,
 reprint,%
]{revtex4-1}

\usepackage{graphicx}
\usepackage{dcolumn}
\usepackage{bm}
\usepackage{caption}
\usepackage{subcaption}
\usepackage{graphicx}
\usepackage[utf8]{inputenc}
\usepackage[T1]{fontenc}
\usepackage{mathptmx}
\usepackage{etoolbox}
\usepackage[dvipsnames]{xcolor}

\makeatletter
\def\@email#1#2{%
 \endgroup
 \patchcmd{\titleblock@produce}
  {\frontmatter@RRAPformat}
  {\frontmatter@RRAPformat{\produce@RRAP{*#1\href{mailto:#2}{#2}}}\frontmatter@RRAPformat}
  {}{}
}%
\makeatother
\begin{document}

\preprint{AIP/123-QED}

\title[Ultra-thin polymer foil cryogenic window for antiproton deceleration and storage]{Ultra thin polymer foil cryogenic window for antiproton deceleration and storage}
\author{B. M. Latacz}
\email{barbara.latacz@cern.ch}
\affiliation{CERN, Esplanade des Particules 1, 1217 Meyrin, Switzerland}
\affiliation{RIKEN, Ulmer Fundamental Symmetries Laboratory, 2-1 Hirosawa, Wako, Saitama, 351-0198, Japan}

\author{B. P. Arndt}
\affiliation{Max-Planck-Institut f{\"u}r Kernphysik, Saupfercheckweg 1, D-69117, Heidelberg, Germany}
\affiliation{GSI Helmholtzzentrum f{\"u}r Schwerionenforschung GmbH, Planckstraße 1, D-64291 Darmstadt, Germany}
 
\author{J. A. Devlin}%
\affiliation{CERN, Esplanade des Particules 1, 1217 Meyrin, Switzerland}
\affiliation{RIKEN, Ulmer Fundamental Symmetries Laboratory, 2-1 Hirosawa, Wako, Saitama, 351-0198, Japan}

\author{S. R. Erlewein}
\affiliation{RIKEN, Ulmer Fundamental Symmetries Laboratory, 2-1 Hirosawa, Wako, Saitama, 351-0198, Japan}
\affiliation{Max-Planck-Institut f{\"u}r Kernphysik, Saupfercheckweg 1, D-69117, Heidelberg, Germany}

\author{M. Fleck}
\affiliation{RIKEN, Ulmer Fundamental Symmetries Laboratory, 2-1 Hirosawa, Wako, Saitama, 351-0198, Japan}
\affiliation{Graduate School of Arts and Sciences, University of Tokyo, 3-8-1 Komaba, Meguro, Tokyo 153-0041, Japan}

\author{J. I. J{\"a}ger}
\affiliation{CERN, Esplanade des Particules 1, 1217 Meyrin, Switzerland}
\affiliation{RIKEN, Ulmer Fundamental Symmetries Laboratory, 2-1 Hirosawa, Wako, Saitama, 351-0198, Japan}
\affiliation{Max-Planck-Institut f{\"u}r Kernphysik, Saupfercheckweg 1, D-69117, Heidelberg, Germany}

\author{P. Micke}
\affiliation{CERN, Esplanade des Particules 1, 1217 Meyrin, Switzerland}
\affiliation{RIKEN, Ulmer Fundamental Symmetries Laboratory, 2-1 Hirosawa, Wako, Saitama, 351-0198, Japan}
\affiliation{Max-Planck-Institut f{\"u}r Kernphysik, Saupfercheckweg 1, D-69117, Heidelberg, Germany}

\author{G. Umbrazunas}
\affiliation{RIKEN, Ulmer Fundamental Symmetries Laboratory, 2-1 Hirosawa, Wako, Saitama, 351-0198, Japan}
\affiliation{Eidgen{\"o}ssische Technische Hochschule Z{\"u}rich, John-von-Neumann-Weg 9, 8093 Z{\"u}rich, Switzerland}

\author{E. Wursten}
\affiliation{RIKEN, Ulmer Fundamental Symmetries Laboratory, 2-1 Hirosawa, Wako, Saitama, 351-0198, Japan}

\author{F. Abbass}
\affiliation{Institut f{\"u}r Physik, Johannes Gutenberg-Universit{\"a}t, Staudinger Weg 7, D-55099 Mainz, Germany}

\author{D. Schweitzer}
\affiliation{Institut f{\"u}r Physik, Johannes Gutenberg-Universit{\"a}t, Staudinger Weg 7, D-55099 Mainz, Germany}

\author{M. Wiesinger}
\affiliation{Max-Planck-Institut f{\"u}r Kernphysik, Saupfercheckweg 1, D-69117, Heidelberg, Germany}

\author{C. Will}
\affiliation{Max-Planck-Institut f{\"u}r Kernphysik, Saupfercheckweg 1, D-69117, Heidelberg, Germany}

\author{H. Yildiz}
\affiliation{Institut f{\"u}r Physik, Johannes Gutenberg-Universit{\"a}t, Staudinger Weg 7, D-55099 Mainz, Germany}

\author{K. Blaum}
\affiliation{Max-Planck-Institut f{\"u}r Kernphysik, Saupfercheckweg 1, D-69117, Heidelberg, Germany}

\author{Y. Matsuda}
\affiliation{Graduate School of Arts and Sciences, University of Tokyo, 3-8-1 Komaba, Meguro, Tokyo 153-0041, Japan}

\author{A. Mooser}
\affiliation{Max-Planck-Institut f{\"u}r Kernphysik, Saupfercheckweg 1, D-69117, Heidelberg, Germany}

\author{C. Ospelkaus}
\affiliation{Institut f{\"u}r Quantenoptik, Leibniz Universit{\"a}t, Welfengarten 1, D-30167 Hannover, Germany}
\affiliation{Physikalisch-Technische Bundesanstalt, Bundesallee 100, D-38116 Braunschweig, Germany}

\author{C. Smorra}
\affiliation{RIKEN, Ulmer Fundamental Symmetries Laboratory, 2-1 Hirosawa, Wako, Saitama, 351-0198, Japan}
\affiliation{Institut f{\"u}r Physik, Johannes Gutenberg-Universit{\"a}t, Staudinger Weg 7, D-55099 Mainz, Germany}%

\author{A. S{\'o}t{\'e}r}
\affiliation{Eidgen{\"o}ssische Technische Hochschule Z{\"u}rich, John-von-Neumann-Weg 9, 8093 Z{\"u}rich, Switzerland}

\author{W. Quint}
\affiliation{GSI-Helmholtzzentrum f{\"u}r Schwerionenforschung GmbH, Planckstraße 1, D-64291 Darmstadt, Germany}

\author{J. Walz}
\affiliation{Institut f{\"u}r Physik, Johannes Gutenberg-Universit{\"a}t, Staudinger Weg 7, D-55099 Mainz, Germany}
\affiliation{Helmholtz-Institut Mainz, Johannes Gutenberg-Universit{\"a}t, Staudingerweg 18, D-55128 Mainz, Germany}

\author{Y. Yamazaki}
\affiliation{RIKEN, Ulmer Fundamental Symmetries Laboratory, 2-1 Hirosawa, Wako, Saitama, 351-0198, Japan}

\author{S. Ulmer}
\affiliation{RIKEN, Ulmer Fundamental Symmetries Laboratory, 2-1 Hirosawa, Wako, Saitama, 351-0198, Japan}
\affiliation{Heinrich Heine University D{\"u}sseldorf, Universit{\"a}tsstrasse 1, D-40225 D{\"u}sseldorf, Germany}

\date{\today}

\begin{abstract}
We present the design and characterisation of a cryogenic window based on an ultra-thin aluminised PET foil at $T<10\,$K, which can withstand a pressure difference larger than $1\,$bar at a leak rate $ < 1\times 10^{-9}\,$mbar$\cdot$l/s. Its thickness of approximately 1.7~$\mu$m makes it transparent to various types of particles over a broad energy range. To optimise the transfer of 100$\,$keV antiprotons through the window, we tested the degrading properties of different aluminium coated PET foils of thicknesses between 900$\,$nm and 2160$\,$nm, concluding that 1760$\,$nm foil decelerates antiprotons to an average energy of 5 keV.  We have also explicitly studied the permeation as a function of coating thickness and temperature, and have performed extensive thermal and mechanical endurance and stress tests. Our final design integrated into the experiment has an effective open surface consisting of 7 holes with 1 mm diameter and will transmit up to 2.5$\,$\% of the injected 100$\,$keV antiproton beam delivered by the AD/ELENA-facility of CERN. 

\end{abstract}

\maketitle

\section{\label{sec:level1}Introduction }

In many areas of physics, chemistry, or engineering, there is a need to separate ultra-high vacuum from high-pressure regions, while nevertheless allowing the transfer of particles between the two sectors. Examples are physics experiments which use gaseous or liquid targets \cite{soter2022high}, plasma-based particle acceleration \cite{tajima1979laser, chen1985, ASCHIKHIN2016175}, trapping of antiprotons \cite{Smorra2015BASEExperiment,smorra2023base,gabrielse2002stacking, kuroda2005confinement,amole2014alpha,amole2015alpha,scampoli2014aegis}, ion cooling \cite{kellerbauer2001buffer, droese2014cryogenic}, or bio-mechanical applications, where control over the types of particles which go through the membrane is required \cite{ULBRICHT20062217, crowley2005isolation, ehrenhofer2016permeation}. The goal of separating high and low pressure sectors is usually achieved either through differential pumping, or by using thin vacuum windows transparent for the incident particles. The main advantage of the differential pumping is that the emittance of the particle beam can be conserved, however, this approach usually requires considerable technical effort and the achievable pressure difference is quite limited. In contrast, using vacuum windows transparent for particles in a certain energy spectrum provides much better vacuum conditions and is technically more efficient to implement, however, this technology causes significant particle beam distortions and limits the available particle flux. \\
In the Baryon Antibaryon Symmetry experiment (BASE) \cite{Smorra2015BASEExperiment}, conducted at the Antiproton Decelerator (AD) and Extra Low ENergy Antiproton ring (ELENA) antimatter facility of CERN \cite{maury1997antiproton}, Geneva, Switzerland, we use ultra-sensitive Penning-trap techniques to study the fundamental properties of single antiprotons, protons, and hydrogen ions, to test the fundamental charge, parity, time reversal (CPT) invariance. These state-of-the-art tests include the measurement of the antiproton magnetic moment with 1.5$\,$p.p.b. (parts per billion) precision \cite{BASE_antip_mag_moment}, and the comparison of the antiproton-to-proton charge-to-mass ratio with a fractional precision of 16$\,$p.p.t. \cite{borchert202216}, constituting the most precise test of CPT invariance in the baryon sector. With this measurement we also conducted the first differential clock-based test of the weak equivalence principle with antimatter. To perform these single (anti)particle experiments, it is essential to store and non-destructively observe antimatter for many months \cite{sellner2017improved}, which requires vacuum pressures below $10^{-17}\,$mbar, constrained by antiproton/residual-gas annihilation cross sections \cite{Fei}. 
The high detection sensitivity of the experiment and the low particle consumption rate means that only ~100 antiprotons need to be transmitted through the window. On the other hand, the vacuum requirements of our non-destructive antimatter experiments are of outstanding challenge.\\ 
In this manuscript, we present a solution to this application with the development, design study, and characterisation of such a vacuum window that is based on aluminised polymer foil. The studied foils have thicknesses between 0.9$\,\mu$m and 2.3$\,\mu$m, and can withstand a pressure difference of up to 1300$\,$mbar with a leak rate below $1\times 10^{-9}$ mbar$\cdot$l/s at temperatures below $10\,$K. We first present the technical design of the window together with the experimental setup. This is followed by a general experimental study of the window leak rate for different parameters such as the foil thickness and the open area of the developed window. In this context we report the first dedicated study of the permeation constant of polymer foils with different aluminium coating thicknesses in room temperature and cryogenic environments and summarise our results of thermal and mechanical endurance  and stress tests. Next, we present the first characterisation of the degrading properties of pure or aluminised PET foils of different thicknesses for a 100$\,$keV antiproton beam, essential information for several other experiments in the field. We conclude our work with the successful demonstration of proton trapping for $296\,$days in a vacuum created by the  degrader window. 

\section{BASE Experimental Setup}

\begin{figure}[htbp]
		\centerline{\includegraphics[width=0.49\textwidth]{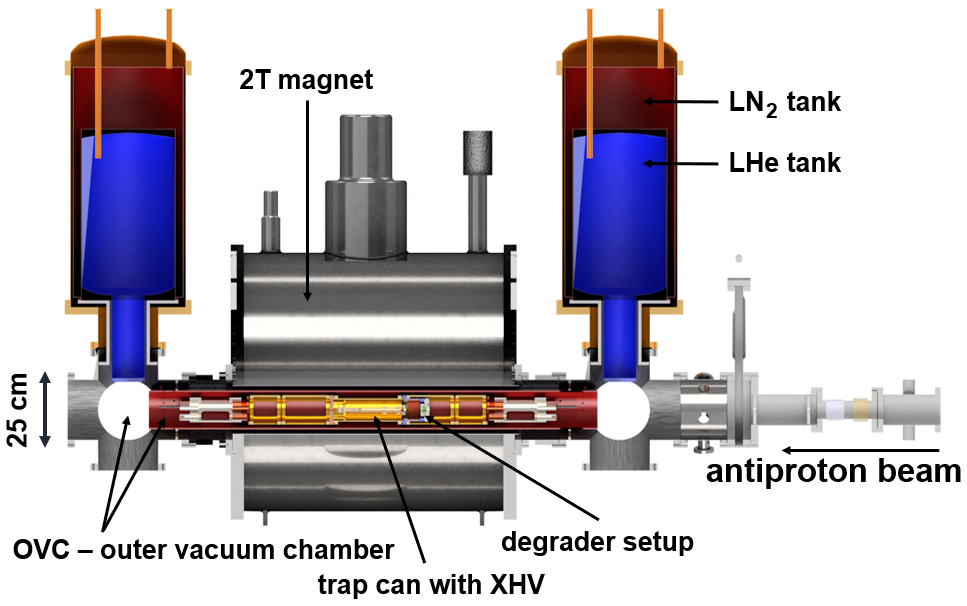}}
		\caption{Cutaway view of the BASE experiment. The Penning-trap electrodes are enclosed inside the trap can which reaches $~10^{-18}$ mbar(XHV). It is placed inside the 2 T superconducting magnet and outer vacuum chamber (OVC), which allows to cool the experiment using liquid nitrogen (LN$_{2}$) and liquid helium (LHe). The cryogenic window (degrader setup) presented in this paper is an interface piece which closes the vacuum of the trap can while allowing the transmission of 100 keV antiprotons to the trap.}
		\label{BASE_general_scheme}
	\end{figure}

An overview of the BASE experimental setup is shown in Fig$.\,$\ref{BASE_general_scheme}. The Penning-trap experiment makes use of a superconducting magnet operated at a magnetic field of $B_0\approx 1.95\,$T with a horizontal bore. The trap itself is operated under cryogenic conditions, the cold temperatures are provided by helium-bath cryostats located on the upstream and the downstream side of the experiment. The upstream side of the apparatus is connected to the AD/ELENA antiproton beamline. Ion-optical elements steer and focus the antiprotons to the entrance flange of the magnet bore and the trap chamber. The outer vacuum in the bore of the magnet, the vacuum crosses that support the cryostats, and the beam tubes, form a closed outer vacuum chamber (OVC), in which pressures of $<10^{-7}\,$mbar are achieved. In the homogeneous center of the magnet the trap is located, enclosed by the \emph{trap can}, which is a cylindrical chamber with a volume of about 1.5$\,$l, made out of high purity (OFHC) copper. 
To separate the trap chamber vacuum from the insulation vacuum, the trap can is closed with custom-made flanges, a feedthrough flange (\textit{pinbase}) on the downstream, and a \emph{degrader flange} on the upstream side. On this upstream degrader flange, described in detail later in the text, a hard-soldered high-purity annealed copper pinch-off tube and the vacuum window are located.\\
The extreme high vacuum (XHV) inside the trap can is achieved by several steps. First, the trap system is mounted in the trap can and closed by the respective flanges. Afterwards, while baking to about 340$\,$K, the closed trap can is pumped via the pinch-off tube to the level of $10^{-7}\,$mbar. Having reached this pressure, the pinch-off tube is sealed by cold-welding it with an appropriate tool. Only after this step the trap is placed in the OVC, which means that the degrader window have to hold an atmospheric pressure. To reach the $10^{-17}\,$mbar pressure required to store antiprotons for at least a year, we accept a maximum leak rate through the window of $\approx1\times 10^{-8}$ mbar$\cdot$l/s for an atmospheric pressure difference at room temperature. In this case, when cooled to cryogenic temperatures, the adsorption of the residual gas on the walls of the trap can and the trap electrodes (cryopumping) is sufficient to reach the XHV pressures required in our experiments.\\ 
The 100$\,$keV energy of the antiproton beam provided by CERN's newly implemented ELENA decelerator \cite{kuchler2014extra} puts another demanding limit on the vacuum window. The degrader must decrease the energy of the antiproton beam from 100$\,$keV to about 1~keV to 5~keV, to allow particles to be captured efficiently in the Penning trap. This necessitates a polymer foil of thickness below 2$\,\mu$m. Simultaneously meeting the demanding requirements of low leak rate, efficient antiproton degrading at acceptable particle flux, robustness, reproducibility and high reliability over months of particle impact and cryogenic operation is a significant technological challenge and the motivation for this work.

\section{Degrader Window}
The main parameter characterising the vacuum, diffusion, and permeation properties of a foil of thickness $d$ and open surface $A$ is its leak rate for a given pressure difference $\Delta p$ equal to
\[ 
L_\text{foil}(A,d,\Delta p) = K_\text{foil} \times \frac{A}{d} \times \Delta p.
\]
\noindent 
Here $K_\text{foil}$ is the permeation constant\cite{engel2020polymer,mapes1994permeation} of the foil.
The vacuum window described in this manuscript is based on biaxially oriented polyethylene terephthalate, H$_{8}$C$_{10}$O$_{4}$, commonly known as Mylar\textregistered ~(DuPont Teijin Films trademark). This material has a permeation constant at the level of $10^{-12}$ m$^2$/s measured at room temperature, and its high mechanical and thermal endurance properties were already reported earlier\cite{HASSENZAHL1975627,engel2020polymer}. 

\begin{figure}
     \centering
       \begin{subfigure}[b]{0.45\textwidth}
         \includegraphics[width=0.99\textwidth]{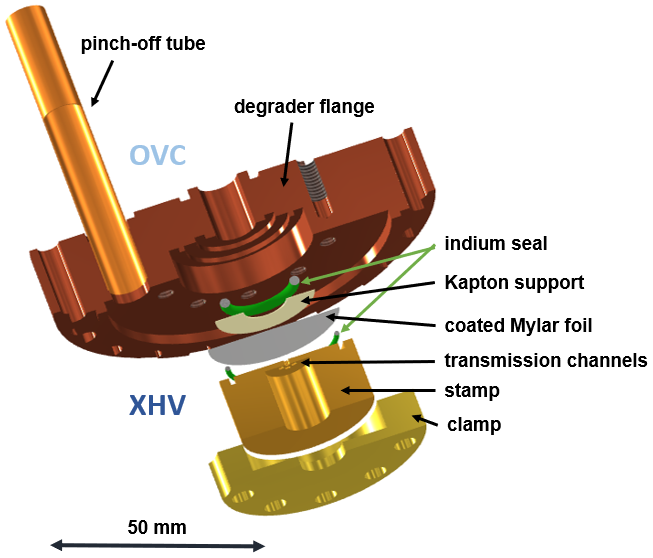}
        \begin{flushleft} (a)
       \end{flushleft}
     \end{subfigure}
    \begin{subfigure}[b]{0.45\textwidth}
         \includegraphics[width=0.99\textwidth]{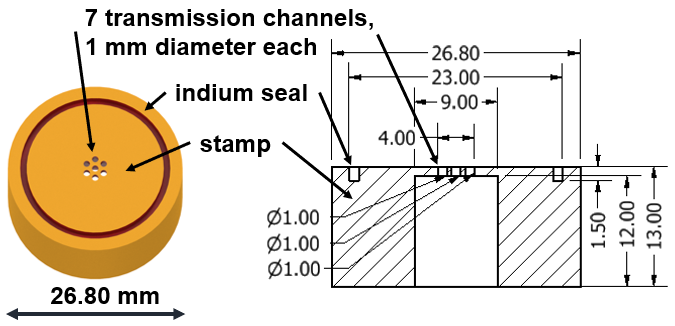}
       \begin{flushleft} (b)
       \end{flushleft}
     \end{subfigure}
  \caption{(a) Detailed scheme of the cryogenic, vacuum-tight degrading window which separates the XHV ($~10^{-18}$ mbar) of the experimental chamber from the surrounding OVC vacuum ($~10^{-7}$ mbar). The main part is the degrader flange with a soldered pinch-off tube which closes the trap can. The seal consists of a tightly pressed indium ring that is placed between the flange and Kapton ring supporting the coated Mylar foil on a stamp, which is kept in place using a clamp. (b) Zoom on the stamp geometry and the transmission channels. On the left an overview drawing and on the right the cross-sectional view with dimensions in mm is shown.}  
  \label{2022-01-17-seal_with_names}
\end{figure}
A detailed schematic which illustrates the technical implementation of the degrader window is shown in Figure \ref{2022-01-17-seal_with_names}. The critical vacuum interface, the \textit{degrader flange}, closes the trap can and separates the XHV of the trap can from the outer isolation vacuum chamber of the experiment. The flange is also  the support structure of the developed semi-transparent vacuum window. Moreover, the pinch-off tube for pre-pumping of the trap-can vacuum is hard-soldered into this flange. 

The window itself is based on \textit{aluminised Mylar foil} glued with a thin layer of Apiezon\textregistered ~N grease \cite{ApiezonN} on the mesh-like \textit{stamp} with several transmission channels that define the final geometrical particle beam acceptance. Both pieces are attached to the degrader flange by a \textit{clamp}, which ensures even pressure distribution over the used seals. As both Mylar and aluminium are not self-sealing, the main seal is achieved using indium placed inside the grooves of the degrader flange, which is squeezed by the stamp with the clamp. The double indium seal is used for additional safety. To protect the foil from being torn by the ductile indium seal, the outer part of the foil is covered with a polished Kapton\textregistered~ring. The sizes of all elements were optimised to fit into the magnet bore of the BASE apparatus \cite{smorra2015base}. The effective transparent surface of the window is defined by the holes in the stamp, which form transmission channels for the particles, see Fig.~\ref{2022-01-17-seal_with_names}(b). To reach the highest possible beam transmission, we performed different systematic studies of various geometries described later in the text. The final configuration used for the experiments with the antiproton beam has a stamp with 7 holes of diameter of 1$\,$mm each, placed in the center and the corners of a regular hexagon with a side length of 1.5$\,$mm. This configuration gives 17$\,$\% geometrical acceptance for the ELENA antiproton beam, which in a plane perpendicular to the beam direction is characterised by a full width at half maximum of about $5\,$mm. Based on empirical studies, a mechanically polished fillet radius of at least 0.1$\,$mm for each hole reproducibly prevents rupture and damage of the foil. Equally important is to polish each part of the seal. This creates uniform surfaces and prevents surface leaks. The stamp implemented into the experiment has a surface roughness \cite{ruvzbarskycontactless} of arithmetic mean deviation $R_{a}=0.42~\mu$m, maximum height $R_{z}=2.17~\mu$m, maximum profile peak height $R_{p}=1.16~\mu$m, mean spacing of profile irregularities $S_{m}=31.12~\mu$m, reduced peak height $R_{pk}=0.25~\mu$m, and reduced valley height $R_{vk}=0.32~\mu$m.

\section{Permeation Constant Measurements}

\begin{figure}
     \centering
       \begin{subfigure}[b]{0.45\textwidth}
         \includegraphics[width=0.99\textwidth]{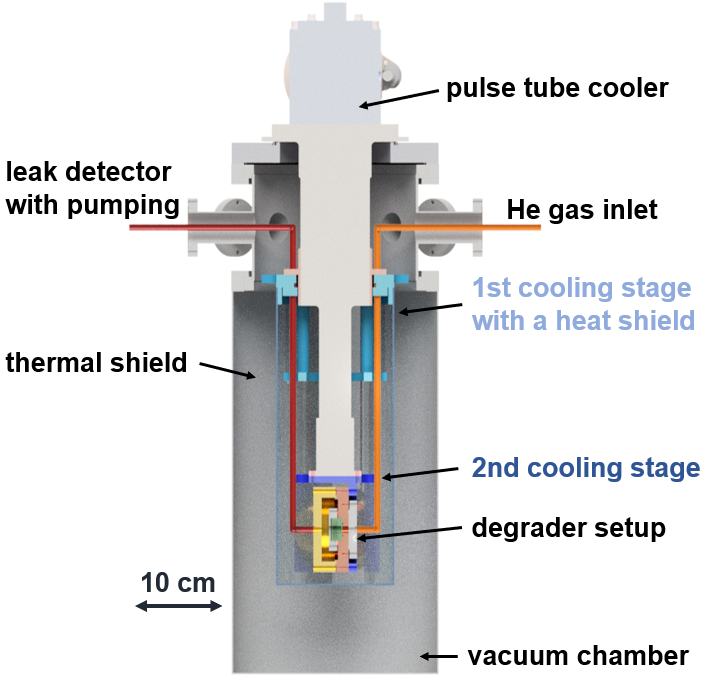}
        \begin{flushleft} (a)
       \end{flushleft}
     \end{subfigure}
    \begin{subfigure}[b]{0.45\textwidth}
         \includegraphics[width=0.99\textwidth]{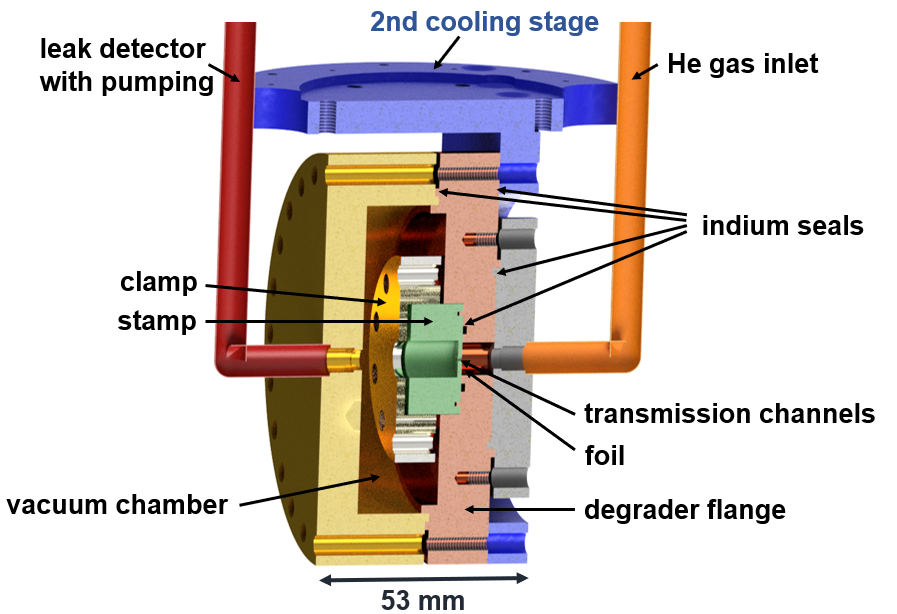}
       \begin{flushleft} (b)
       \end{flushleft}
     \end{subfigure}
  \caption{Schemes of the setup used to test the cryogenic window at temperatures down to 2.7 K. (a) A full scheme of the setup based on the Sumitomo RDK-408D2 cryocooler with two cooling stages. The helium gas is supplied to the test chamber via an 8 mm diameter tube from the gas bottle. The pressure applied to the system is controlled in multiple stages using a pressure reducer and helium preparation vacuum chambers. The other side of the test chamber is pumped via a leak detector which allows a measurement of the flow of the helium gas through the tested cryogenic window. (b) Zoom of the 2nd (2.7 K) cooling stage where the chamber with the degrader is placed.}
  \label{Cryocooler_test_setup}
\end{figure}

During the window development phase we performed different measurements of the helium leak rate through various foils both at room- and cryogenic temperatures, using the experimental setup shown in Fig$.$~\ref{Cryocooler_test_setup}. The setup consists of a degrader chamber which can be cooled to $T<10\,$K using a Sumitomo RDK-408D2 cryocooler. All elements of the degrader chamber connected to the second cooling stage are made of oxygen-free high thermal conductivity (OFHC) copper for the best possible thermal conductivity and nonmagnetic properties. Using Swagelock stainless steel connections, one side of the degrader chamber is pumped using the turbomolecular pump built into a Leybold Phoenix L300i leak detector, which allows us to measure the helium leak rate through the designed window down to the level of $1\times10^{-12}$ mbar$\cdot$l/s. The other side of the degrader chamber is connected to the helium supply, which provides up to 2 bars of He pressure. We use He as a test gas as it has the highest permeation constant. Before inserting gas, the chamber is first pumped to a pressure around $1\times 10^{-1}$ mbar.

In this section, we characterise the leak rate through different foils. Unless otherwise noted, the measurement was performed for 1$\,$bar of helium pressure applied to the inlet, measured using a Baratron 120AA from MKS Instruments. To establish that any measured leaks were due to permeation through the foils rather than leaks through the indium seals, we performed different leak rate measurements with a 200 $\mu$m thin aluminium foil instead of the Mylar foil. Both at 300$\,$K and at cryogenic temperatures, these measurements reached the background sensitivity of the leak detector, from which we conclude that the measured leak rate corresponds to the permeation through the degrader foil itself. This was also confirmed by executing reproducibility tests using the same Mylar foil in technically identical assemblies but in different characterisation runs.

\subsection{Tested Materials}

We tested a variety of pure Mylar foils with thicknesses between $d=0.5\,\mu$m to $d=2.5\,\mu$m, and Mylar foils metallised on one or both sides and varying thickness of the aluminium layers. The aluminium layer was coated using two methods, either magnetron sputtering  \cite{gruen1991process} or evaporation of metal in vacuum \cite{klauk2006organic}. Magnetron sputtering allows the production of aluminium coating thicknesses between 25$\,$nm and 200$\,$nm, as stated by the producer of the foil and confirmed by measurements using a precise weighting scale. We obtained Al-coating thicknesses of 30$\,$nm and 80$\,$nm for the evaporation-coated foils, as these were commercially available thicknesses.

\subsection{Impact of the aluminium coating}

\begin{figure}[htbp]
		\centerline{\includegraphics[width=0.45\textwidth]{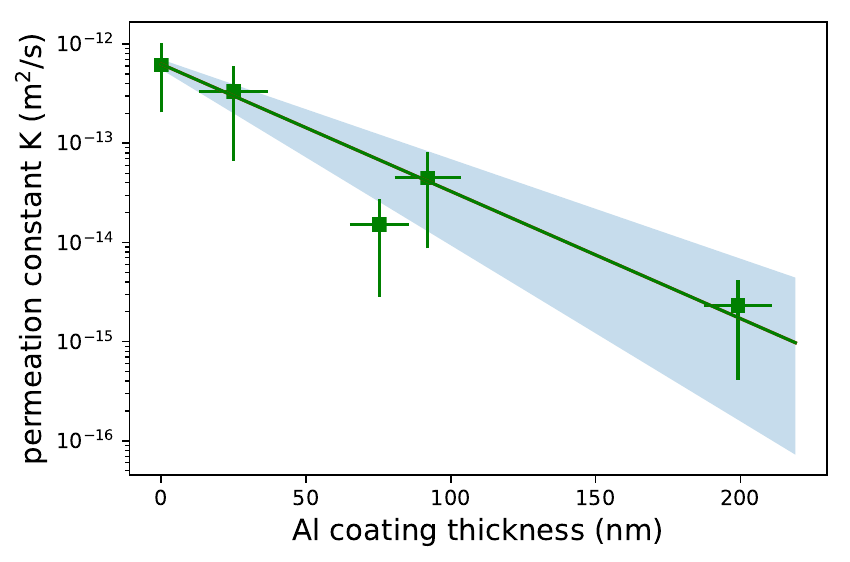}}
		\caption{Permeation constant for 900 nm thick Mylar foil as a function of the aluminium coating thickness added on both sides together with a shaded $\pm$2$\sigma$ confidence interval of the fitted exponential function. Uncertainties are dominated by the measurement of the foil thickness.
		Each leak measurement is an average of over 60 min of data (3600 points), collected 60 minutes after start of pumping. These measurements were performed with a stamp with 7 1-mm-diameter holes. The point at 0~nm aluminium thickness is a reference point to a pure Mylar foil measured with a stamp with one 1.2 mm diameter hole. Measurement performed at room temperature.}
		\label{leak rate_Al_coating_thickness}
	\end{figure}

Permeation through biaxially oriented polymer foils can be suppressed by metallising the polymer film. The coating closes the porous polymer structure and suppresses atomic and molecular diffusion. To investigate this effect quantitatively, we study the leak rate as a function of coated Al thickness $\lambda$. As a base material, 900 nm Mylar foil coated on both sides was used. Four different magnetron sputtered foils with coating thicknesses between 25$\,$nm and 200$\,$nm were tested, and the measured permeation constants are plotted in Fig.~\ref{leak rate_Al_coating_thickness}.

Using the experimental setup described above and mounting the foils with the 7-hole support stamp to close the window, the leak rate of each of the foils was measured at 300$\,$K. In the experiments the system was first pumped for 60$\,$min, afterwards the high pressure side of the window was vented with 1$\,$bar of helium, and the leak rate was recorded for 60$\,$min with a 1$\,$s sampling rate. The mean results of this characterisation campaign are shown in Fig.~\ref{leak rate_Al_coating_thickness}. The quoted uncertainties reflect the fluctuations of the measurements and the shaded area covers the $\pm$2$\sigma$ confidence interval of a fit with an exponential function of type $\propto \exp(-\lambda/ \lambda_{0})$ with the effective permeation reaching $\lambda_{0} = 33.9 \pm 4.7$ nm. The point at zero aluminium thickness is a reference point using an uncoated Mylar foil. We observe that the change in the thickness of the aluminium layer from 25$\,$nm to 200$\,$nm decreases the leak rate by two orders of magnitude. We note that the mechanical properties of the foil with 200$\,$nm Al-coating were dominated by the properties of the fragile aluminium-layer, prone to cracking when stretched.

Results of less systematic but similar experiments with foils produced by vapor metallization are summarized in Table I, showing qualitatively similar behaviour for these foils.

\begin{table}[h!]
\centerline{
\begin{tabular}{ c|c|c } 
 \hline
 Foil & $K_{\text{foil}}$ [$m^{2}/s$]  & Leak rate [mbar$\cdot$l/s] \\ \hline
2.5 $\mu$m Mylar & $1 \times  10^{-12}$ & $5\times 10^{-7}$ \\ \hline 
2 $\mu$m Mylar+80$\,$nm Al & $4 \times  10^{-14}$ & $2.2\cdot 10^{-8}$ \\ \hline
2 $\mu$m Mylar+80$\,$nm Al on each side & $4.6\times 10^{-15}$ & $1.9 \cdot 10^{-9}$\\ \hline
\end{tabular}
}
\label{table1}
\caption{Permeation constants of Mylar foils without, with one, or both sides aluminised (layer thickness 80$\,$nm). The foils were metallised by evaporation. Measured at room temperature with one 1.2 mm diameter transmission channel.}
\end{table}


\subsection{Optimisation of the stamp geometry}

Aluminised Mylar foil in the thickness range as investigated here is very sensitive to stretching and sharp edges. That is why a circular shape of the window is used so that the forces introduced by stretching and cooling are uniformly distributed over the foil surface. To optimise the size of the window, we studied the leak rate through a 2$\,\mu$m Mylar foil coated on both sides with 80$\,$nm aluminium (later referred to as 2160$\,$nm aluminised Mylar foil), while supporting it with a stamp with a single central hole. In these experiments we varied the hole diameter and determined the permeation constant of each assembly. The results as a function of the single hole diameter are presented in Table II. In the range between 1$\,$mm and 1.5$\,$mm hole diameter we measure permeation constants that are similar within the measurement uncertainties. At 2$\,$mm diameter, the permeation constant of the foil was observed to continuously increases due to developing damage of the foil structure caused by stretching. For our geometries this behaviour was reproducibly observed in repetitive experiments. On the other hand, up to a 1.5$\,$mm diameter hole the leak rate was constant even after many days of data sampling. 

\begin{table}[h!]
\centerline{
\begin{tabular}{ c| c } 
 \hline
Hole diameter (mm) & Permeation constant K (m$^2$/s) \\ \hline
1.0 & $3.4(1.3) \times 10^{-15}$ \\ \hline
1.2 & $4.6(1.5) \times 10^{-15}$ \\ \hline
1.5 & $3.7(1.1) \times 10^{-15}$ \\ \hline
\end{tabular}
}
\label{foil_aluminium_coating_table_summary}
\caption{Measured leak rate through the 2160 nm aluminised Mylar foil as a function of a single hole diameter.}
\end{table}

\begin{figure}[htbp]
		\centerline{\includegraphics[width=0.45\textwidth]{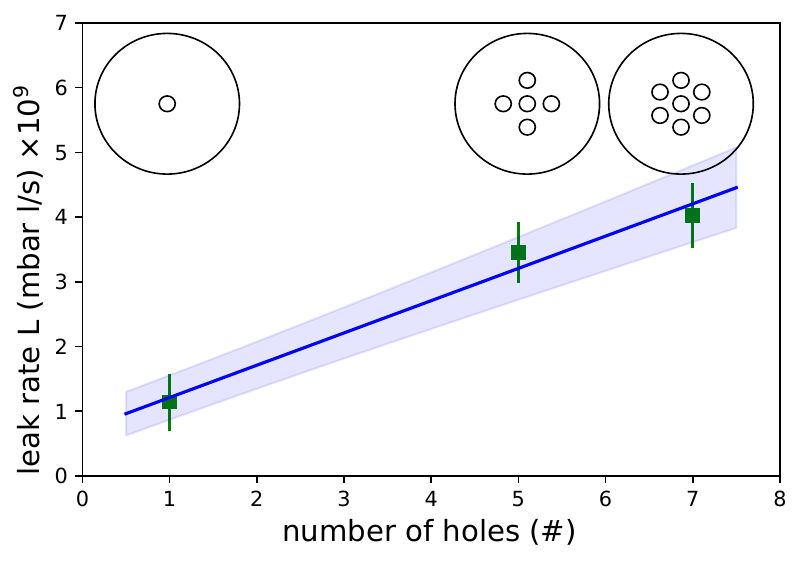}}
		\caption{Measured leak rate through the 2160 nm aluminised Mylar as a function of the number of 1-mm-diameter holes with a fitted straight line (blue) together with a shaded 2$\sigma$ confidence interval.}
		\label{Al2M20_number_of_holes}
\end{figure}

Selecting a hole diameter of 1$\,$mm, we have investigated the leak rate as a function of the number of holes distributed over the surface of the degrader window, the results of this study are shown in Fig$.\,$\ref{Al2M20_number_of_holes}.

Up to seven holes we see a linear scaling of the leak rate as a function of transparent surface area, indicating no inelastic damage while pumping the foil. For an even higher number of holes and the selected geometries we observed irreversible damage. In the final design we chhose to use 7 holes, as the transparency of the window with the centered hexagonal hole distribution with respect to the ELENA beam is at 17$\,\%$.

\section{Cryogenic Characterisation}

\begin{figure}[htbp]
\centerline{\
		\includegraphics[width=0.49\textwidth]{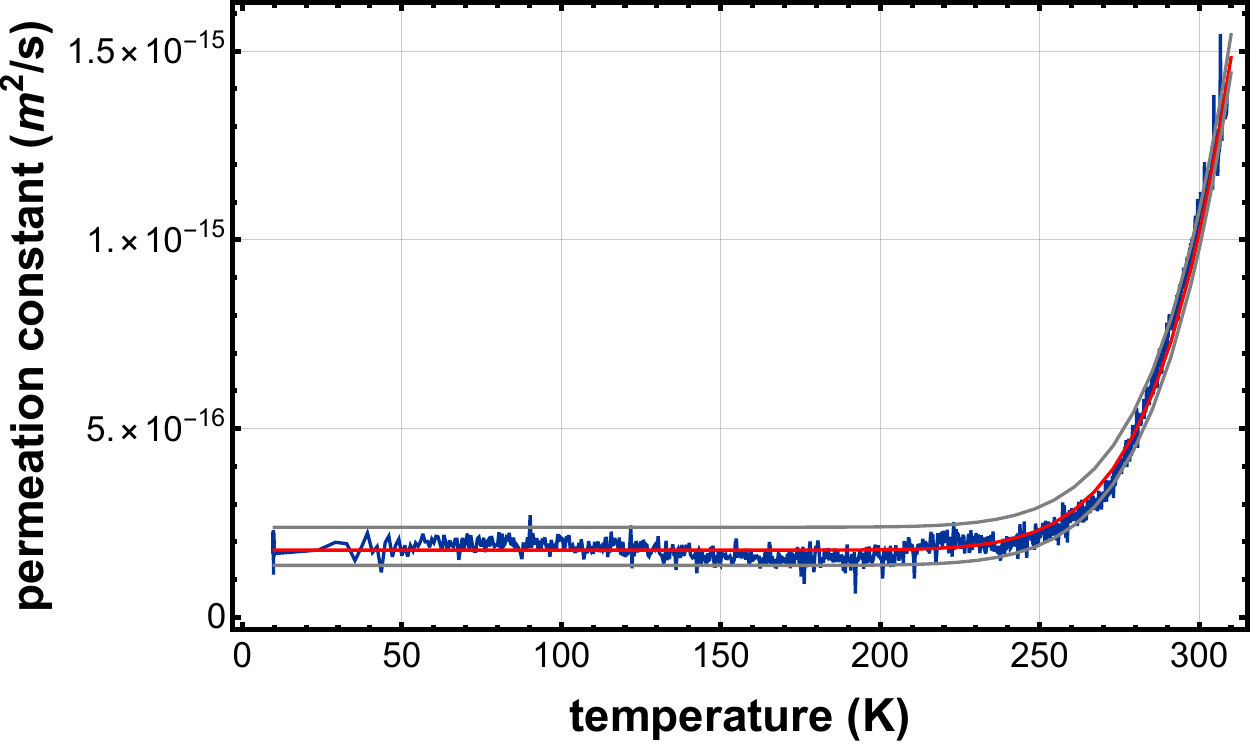}}
		\caption{Leak rate through a 1960 nm thick foil as a function of its temperature. Data were treated with a median filter with a time step of 19.8 s.}
		\label{1960_heating_curve}
\end{figure}

In the BASE experiment, the described window will be operated at a temperature of about 4.5$\,$K, to separate the XHV of the trap can from the isolation vacuum of the experiment. That is why for foils of thicknesses 2160$\,$nm, 1960$\,$nm, and 1760$\,$nm, all coated on both sides with a vaporized Al-layer of 30$\,$nm thickness, we performed various endurance tests. Within this testing campaign, we exposed the foil assemblies to mechanical stress under room temperature- and cryogenic conditions, while measuring the leak rate with the cryogenic test-setup described above (see Fig.~\ref{Cryocooler_test_setup}). Additionally, using a similar experimental setup, we expose the windows to thermal shocks, by pouring a direct stream of liquid nitrogen onto the foils while continuously measuring the leak rate.\\
The summary of the measured average leak rate for the three tested foils at room temperature and at cryogenic temperatures below 10$\,$K are presented in Table III. We observe that the leak rates at $T=300\,$K and $T<10\,$K differ by about one order of magnitude.\\
The temperature dependence of the permeation constant through polymers can be empirically described by the van Hoff-Arrhenius relation \cite{fuoco2018temperature}, where the dependence is parameterized by the effective sorption, desorption and diffusion dynamics that obey effective thermodynamic scaling laws $\propto\exp\left(-\Lambda_p/T\right)$. The leak rate as a function of temperature was measured while the apparatus was warming up, exemplary data shown in Fig.~\ref{1960_heating_curve}. Interestingly the leak rate as a function of temperature reproducibly shows for all the tested foils a rapid decrease in the temperature range between 300$\,$K and 250$\,$K, then staying constant down to $T<10\,$K.

\begin{table*}[htb]
\centerline{
\begin{tabular}{ c| c |c} 
 \hline
 Foil & Average leak rate at 293 K (mbar$\cdot$l/s)  &  Average leak rate at 6-10 K (mbar$\cdot$l/s) \\ \hline

2 $\mu$m Mylar + 80 nm Al on both sides & $5.49(0.02) \cdot 10^{-9}$ & $2.49(0.09) \times 10^{-11}$  \\ \hline
1.9 $\mu$m Mylar + 30 nm Al on both sides & $1.467(0.004) \times 10^{-8}$ & $4.8(0.6) \times 10^{-10}$  \\ \hline
1.7 $\mu$m Mylar + 30 nm Al on both sides & $1.46(0.07) \times 10^{-8}$  & $6.57(0.05) \times 10^{-10}$ \\ \hline
\end{tabular}
}
\label{foil_aluminium_coating_table_summary_2}
\caption{The average leak rate measured for different foils at both room and cryogenic (below 10 K) temperatures. Each value is an average of 9000 measurement points, covering a time interval of at least 1 hour. The leak was measured under 1200 mbar helium pressure measured at room temperature with a baratron.}
\end{table*}

\begin{figure}[htbp]
\centerline{\
		\includegraphics[width=0.49\textwidth]{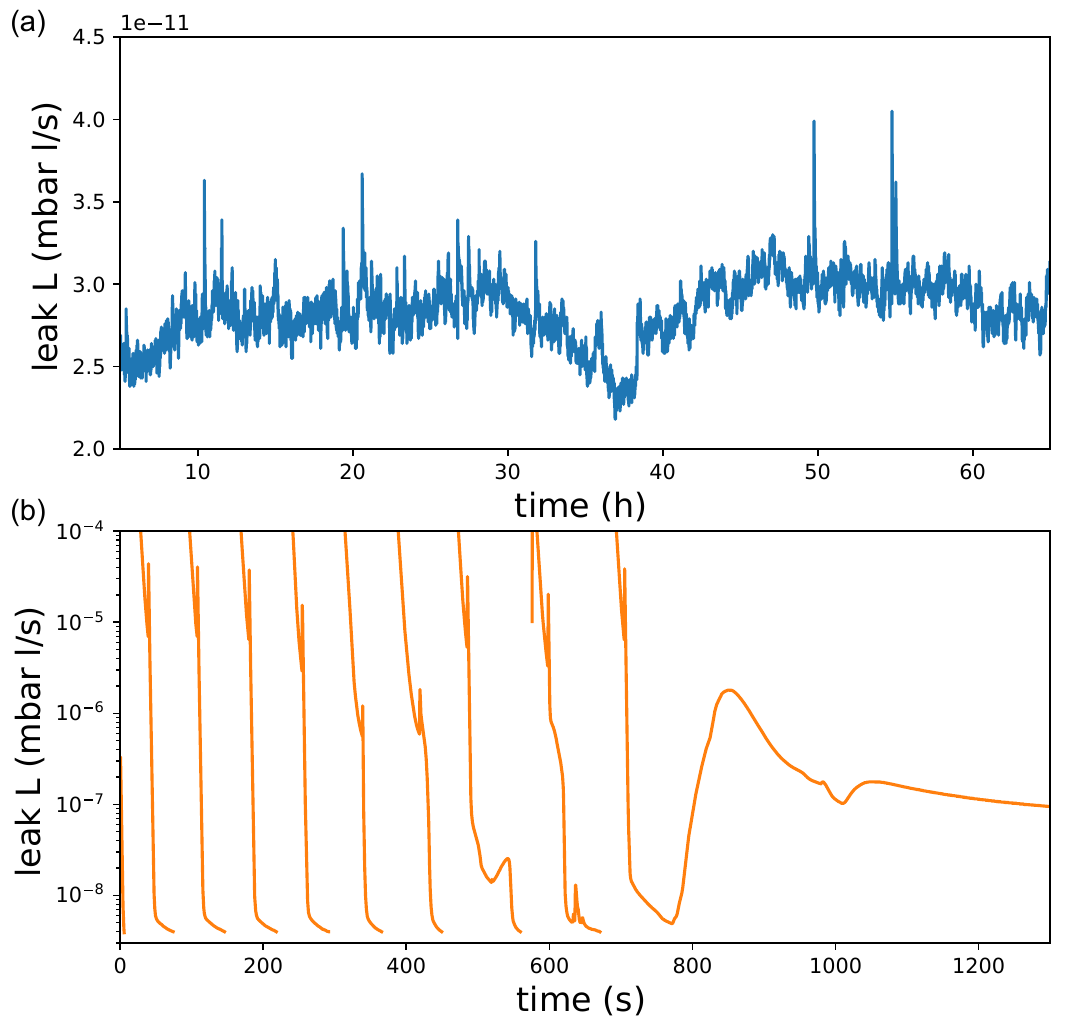}}
		\caption{Cryogenic endurance tests with a 2160 nm thick foil. (a) A 68 hours long exemplary interval of the leak rate measurement at 10 K. The full experiment lasted for 25 days. (b) Leak rate as a function of time, while performing cryogenic endurance test carried out at 77 K with liquid nitrogen. The foil was continuously pumped and re-pressurised. The leak rate was limited by thermal expansion of the support vacuum chamber, for further details we refer to the text. The test was performed with a single hole with 1.2 mm diameter as transmission channel.}
		\label{LN_4K_endurance_tests}
\end{figure}

For the smooth long-term experimental operation in a cryogenic precision experiment, long-term stability, resilience with respect to repetitive cooling/warming cycles and stretching due to a change of pressure difference experienced by the foil is essential. 
To study the long-term stability of the degrader window we kept the 2160$\,$nm thick foil at 10$\,$K temperature under a stationary helium pressure of 1200$\,$mbar for 25 days. Figure \ref{LN_4K_endurance_tests}(a) shows a 68\,h long interval of the measured leak rate, representative of the stability that was observed over the entire 25 days. The 25$\,\%$ baseline drift around the level of $2\times 10^{-11}$ mbar$\cdot$l/s is correlated to laboratory temperature changes, the fluctuation spikes with amplitudes at the $10^{-10}$ mbar$\cdot$l/s level are likely induced by vibration and out-gassing of micro-enclosures in the beam tubes. 
Before and after this measurement, the foil was thermally cycled for several times and exposed to ten sequences of pressure changes between 0.1$\,$mbar and 1200$\,$mbar at ramp times between 10$\,$s and 100$\,$s, and eventually the system was re-cooled and the leak rate was measured again. During of all these stress-tests, in load amplitude and time constants much higher than the expected changes during experiment operation, we did not detect an increase of the measured leak rate within the 5$\,\%$ resolution limit of the measurement.\\
Lastly, while measuring the leak rate, we exposed the window to thermal shocks by sinking it into liquid nitrogen (LN$_{2}$), corresponding to quasi-instantaneous temperature changes of 220$\,$K per cycle. Additionally, we performed stretching tests at LN$_{2}$ temperature, by venting the pumping system and the leak detector, thus changing the pressure difference experienced by the foil. In these tests we pumped the system until the measured leak was smaller than $4\times10^{-9}$ mbar$\cdot$l/s, then vented the system to 1$\,$bar of He pressure and pumped again afterwards. The result of this test is shown in Fig$.\,$\ref{LN_4K_endurance_tests}(b), eight full pumping cycles were successful. The reduced performance after the ninth cycle was found to be caused by a cryo-leak in the supporting vacuum chamber, not by degradation of the window.\\
\\
In summary, all three foils, 2160$\,$nm, 1960$\,$nm and 1760$\,$nm, were cooled at least twice, with multiple foil stretching tests performed by varying the helium pressure in the system both at room and cryogenic temperatures. Within these measurements, we were not able to resolve any indication of degradation or mechanical fatigue behaviour of the foil. The window also withstands long term operation and cryogenic temperature and survived endurance, pressure- and thermal shocks that are partly by three orders of magnitude more extreme than expected under the final experimental conditions. Thus, all three foils meet the requirements to be implemented into the experiment.

\section{Antiproton deceleration}

The vacuum window presented above has the crucial role of being a degrader for the 100 keV antiproton beam delivered by CERN's ELENA decelerator. In this section we present the results of the characterisation of the degrading properties of Mylar foils of different thicknesses, which allows matching the thickness of the window to the initial and required final energy of antiprotons. 

\subsection{Theoretical estimation of the required foil thickness}\label{theory_simple_model}

At the time when the window was designed, there was a lack of experimental data and models for the stopping power of antiprotons in Mylar in the energy range between 1$\,$keV and 100$\,$keV. To approximately estimate the range of thicknesses for the degrading Mylar foil, we developed an effective guiding model  based on the measured stopping power data of antiprotons in carbon \cite{moller2002antiproton}, and the available stopping power data of protons in different materials. The data for protons were taken from the scientific community code SRIM \cite{ziegler2010srim}. According to Bragg's rule \cite{thwaites1983bragg}, to first order, the stopping power $S(E)$ in a compound material can be estimated as a linear combination of the stopping powers of its components. In case of Mylar $\text{H}_{8}\text{C}_{10}\text{O}_{4}$ this results in
\[
S(E)_{\text{Mylar/p}} = \frac{8}{22}\times S(E)_{\text{H/p}}+ \frac{10}{22}\times S(E)_{\text{C/p}} + \frac{4}{22}\times S(E)_{\text{O/p}} 
\]
\noindent 
where $S(E)_{H,C,O/p}$ are the stopping powers of protons in hydrogen, carbon and oxygen, respectively. In more recent calculations that account for atomic bonding effects, Bragg's formula is multiplied by the compound correction, which is for Mylar given as 0.9570 \cite{ziegler2010srim}.
To a good approximation, it is expected that the stopping power for antiprotons in Mylar can be described using the same formula but using stopping powers for antiprotons. 
To compensate for the lack of available  $S(E)_{\bar{p}}$ data for antiprotons, which was in the expected energy range only measured for  carbon or aluminium \cite{moller2002antiproton}, we describe the $S(E)_{Mylar/\bar{p}}$ as $S(E)_{C/\bar{p}}$ multiplied by a scaling factor $\lambda$
\[
S(E)_{\text{Mylar}/\bar{\text{p}}} = \lambda \times S(E)_{\text{C}/\bar{\text{p}}},
\]
\noindent 
where  $S(E)_{C/\bar{p}}$ is an extrapolated fit to the data from Ref.~\onlinecite{moller2002antiproton}.
To obtain a realistic estimate of the value of the effective parameter $\lambda$, we use data for the stopping power of protons in Mylar and carbon and calculate $\lambda$ as
\[
\lambda = \frac{S(E)_{\text{Mylar/p}} }{S(E)_{\text{C/p}}} = 0.620 \pm 0.002
\]
\noindent 
for the energy range between 1$\,$keV and 100$\,$keV. We show below that this effective approach is within the measurement uncertainties in agreement with our measurements.

\begin{figure}[htbp]
\centerline{\
\includegraphics[width=0.5\textwidth]{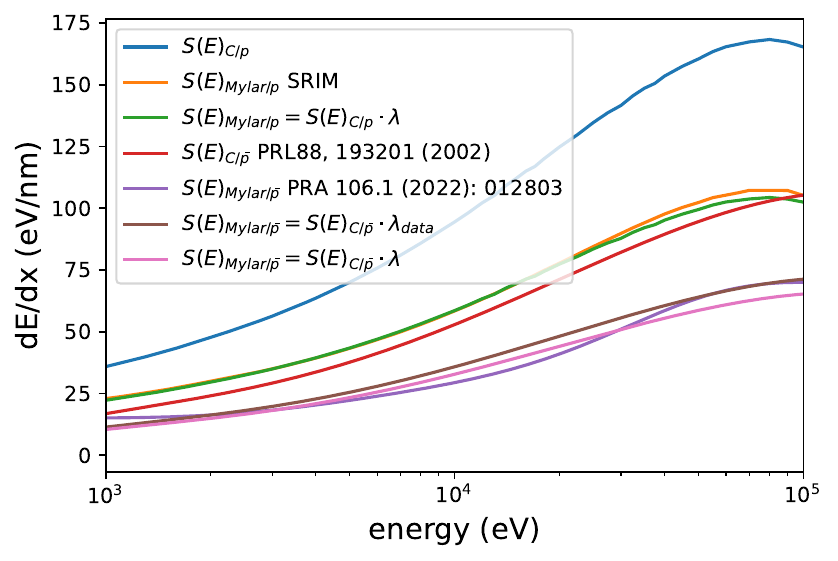}}
    \caption{Stopping power values for protons $\text{p}$ or antiprotons $\bar{\text{p}}$, for: carbon - $S(E)_{C/\text{p}(\bar{\text{p}})}$; Mylar (from SRIM) - $S(E)_{Mylar/\text{p}}$; Mylar (from  Ref. \onlinecite{nordlund2022large}) - $S(E)_{Mylar/\bar{\text{p}}}$; Mylar according to model described in the text - $S(E)_{Mylar/\bar{\text{p}}}\cdot \lambda$; Mylar according to model described in the text with parameter $\lambda$ fitted to the data - $S(E)_{Mylar/\bar{\text{p}}}\cdot \lambda_{data}$.}
\label{stopping_power_plott}
\end{figure}

Recently, the first experimental data for the stopping power of low-energy antiprotons for 1800$\,$nm Mylar foil coated with two 25$\,$nm thick silver layers were published in Ref.~\onlinecite{nordlund2022large}. In the same paper, the authors provide calculations of the electronic and nuclear stopping power of antiprotons in Mylar using a molecular dynamics approach. The main advantage of these results is that for the first time nuclear scattering effects for antiprotons were included in the calculation, which significantly enlarged the values of the nuclear stopping power around 1$\,$keV energies. The comparison of this refined model to our measurements is also shown in the next section. Figure \ref{stopping_power_plott} compares the stopping power curves in an energy range between 1 keV and 100 keV which were described in this Section together with a stopping power curve obtained from the measurements described in the following text. 

\subsection{Degrading properties of the Mylar foil}

Using CERN's 100$\,$keV ELENA beam, we tested the degrading properties of different thicknesses of Mylar or aluminised Mylar foils, to explicitly measure  the optimum foil thickness for antiproton injection into our system. For this, we performed a time-of-flight measurement of particles transmitted through the degrader within a 6$\,^\circ$ angle, the relevant data were recorded with a set of scintillation counters placed along the antiproton transfer line \cite{Smorra2015BASEExperiment}. The transmitted particles annihilated at least 65$\,$cm behind the degrader, which enables us to distinguish them from particles annihilating inside the degrader, and allows for an estimation of the energy of the slowed antiprotons. For a better understanding of the transmitted signal, we used a movable target whose perpendicular position with respect to the beam axis was adjusted using a piezo-driven linear slip-stick stage, specified for operation in high magnetic fields and under ultra-high vacuum conditions \cite{piezo_drive}. The target had three positions: open - in which the beam transmits undisturbed, block - in which the entire beam annihilates on the target, and foil - where the beam transmits through a 9$\,$mm diameter aperture covered by the foil under test. This combination of measurements allows to distinguish different backgrounds originating from beam annihilation in different places of the system. Tests were performed for four different foils: 900$\,$nm thick pure Mylar, 900$\,$nm Mylar with two 100$\,$nm thick layers of Al, 1400$\,$nm, and 1700$\,$nm both foils covered with two Al-layers of 30$\,$nm thickness.

\begin{figure}[htbp]
\centerline{\
		\includegraphics[width=0.5\textwidth]{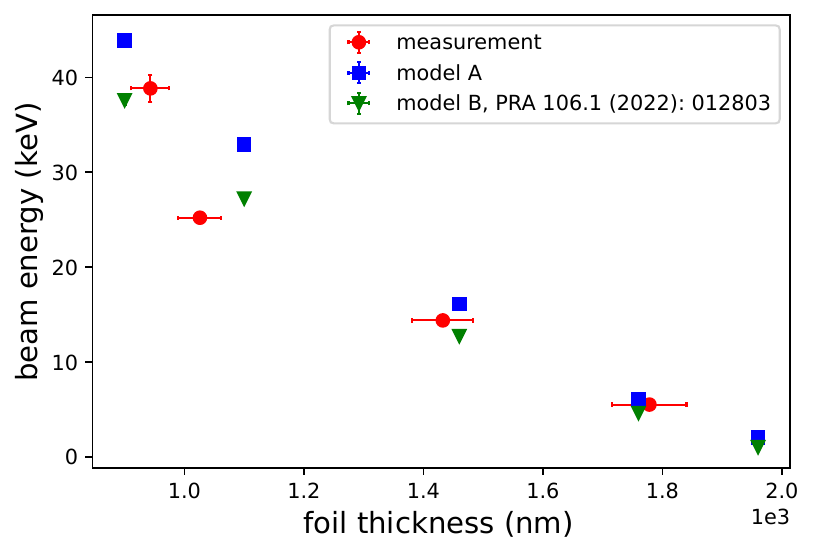}}
		\caption{Measured energy of the transmitted beam as a function of the foil thickness (red round points) compared to the simple phenomenological model presented in Section \ref{theory_simple_model} based on measured data of the stopping power of antiprotons in carbon \cite{moller2002antiproton} (blue squares) and the model presented in Ref. \onlinecite{nordlund2022large} (green triangles).}
		\label{energy_tof}
\end{figure}

The measured energy of the transmitted beam as a function of the foil thickness is shown in Fig.~\ref{energy_tof}. We determine the mean energy based on the simulated pulse-shapes using SRIM \cite{ziegler2010srim}, and obtain the mean energy by deconvolving the data with simulation-based effective profiles, that reproduce the data within the residuals of the background noise of the detector. Small deviations of the expected profiles are attributed to annihilations on obstacles along the beamline of the cryogenic Penning trap experiment. A detailed study of the consistency of model assumptions and measured profiles would require a redesign of parts of the cryogenic Penning trap experiment and are beyond the scope of this manuscript. The data are compared to simple simulation results in which particles go straight through the pure Mylar foil, assuming two different models for the stopping power of Mylar - model A described in Section \ref{theory_simple_model} and model B taken from the Ref.~\onlinecite{nordlund2022large}. Taking into account only particles which go straight through a Mylar foil is a correct approximation as the measured delayed signal originates from particles which are transmitted through the foil with angular transverse momentum spread smaller than 6 degrees. In the simulations, we are neglecting the difference between the stopping power of aluminium and Mylar, as the thickness of aluminium is small relatively to the thickness of Mylar, and the measured stopping power of aluminium \cite{moller2002antiproton} is only up to 30 \% larger than the estimated stopping power for Mylar. The new calculations presented in Ref. \onlinecite{nordlund2022large} confirm this assumption. Our measurements and both models are in qualitative agreement. Based on the fit to the data we extracted the parameter $\lambda_{data} = 0.677\pm0.066$, which is in agreements with our theoretical prediction and can be used to simulate in a simple way the behaviour of antiprotons going through Mylar. 

For efficient trapping, BASE can accept particles between 1 keV and 5 keV energy after the degrading stage. That is why in the final assembly we decided to use foils of thickness 1760 nm and 1960 nm, which should allow to transmit up to 2.5 \% of the injected 100 keV antiprotons.

\section{Integration into the BASE experiment}

\begin{figure}[htbp]
		\centerline{\includegraphics[width=0.5\textwidth]{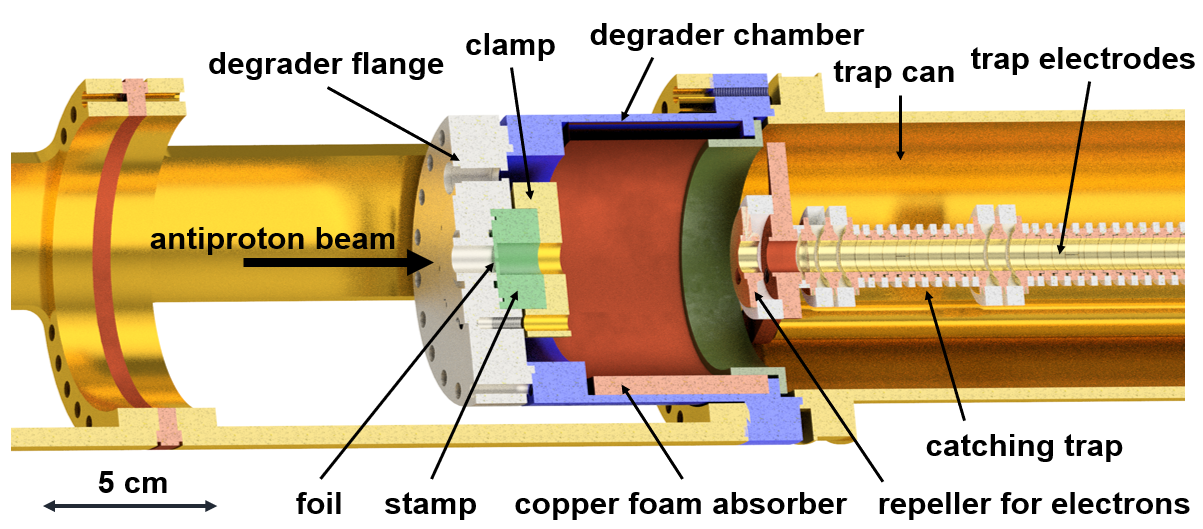}}
		\caption{Scheme of the new, vacuum-tight, micrometer-thick degrader window implemented in the BASE experiment.}
		\label{BASE_experiment_setup}
\end{figure}

The implementation of the designed foil system into the BASE setup, as a degrader for the antiproton beam and as XHV window to close the trap can, is shown in Fig$.~$\ref{BASE_experiment_setup}. While operating the experiment, all components shown in the drawing are at $T<5.2\,$K. The degrader flange is interfaced to the trap can by a \textit{degrader chamber}, in which copper foam absorber is placed to increase the adsorption surface for cryopumping. The copper foam increases the cryogenic metal surface in the trap can by a factor of 2.5. The distance between the degrader window and the center of the most upstream trap, the reservoir or catching trap, is $\approx 11.6\,$cm. The entire volume between degrader window and trap center is placed in a background magnetic field $>1.5\,$T.
We note that upstream to the reservoir/catching trap a repeller electrode is placed. Antiproton catching relies on electron cooling. Before antiprotons are injected, a cloud of electrons is loaded into the catching trap which rapidly cools to thermal background temperatures by the emission of cyclotron radiation. Antiprotons injected into a high-density electrons are sympathetically cooled by scattering on- and cyclotron radiation of electrons. To protect the degrader foil from the 40$\,$eV to 80$\,$eV electron beam, which is typically operated at currents of 20$\,$nA to 200$\,$nA, the repeller electrode is biased to twice the kinetic energy of the electron beam.

\subsection{Trapped charged particles}

For the commissioning of the experiment we load a cloud of protons using molecular hydrogen dissociation mechanism by the electron beam. With the designed vacuum system, closed with the 1960$\,$nm foil, no particle loss or decreased performance of the experiment was observed throughout a continuous observation time lasting from the 10th of October 2021 to the 1st of August 2022, i.e.~for 295 days. We also observed that the implemented vacuum window robustly survived several experiment cooling cycles between 300$\,$K and 5$\,$K, at temperature ramping rates of 24$\,$h per cycle for both warm-up and cool-down, an important feature for the practical operation of the experiment. 
Operating the system under cryogenic and high vacuum conditions for 295$\,$days, the experiment was warmed up for technical radiofrequency maintenance and the degrader window was extracted and investigated again in the experimental test setup. In these studies we measured the leak through the foil at $8.5\times 10^{-8}$ mbar$\cdot$ l/s. Currently it is not possible to conclude whether the slightly increased leak rate from $1.46(0.07)\times10^{-8}$ mbar$\cdot$ l/s (Table III) is related to mechanical fatigue or to over-pressurization in the reassembled test setup.
To further optimize antiproton catching in BASE, the next run was operated with the 1760$\,$nm foil. Up to the date of writing this manuscript (216 days) no decrease in the experiment performance due to increased pressure was observed. The lowest pressure limit that was estimated based on explicit measurements comes from the observation of quantum heating rates of a single trapped proton in the strong magnetic bottle of the analysis trap of the experiment \cite{Smorra2015BASEExperiment}. By initially cooling the magnetron mode to particle energies $<0.6(6)\,\mu$eV, performing experimental campaigns for typically 72$\,$h and recooling the magnetron mode afterwards, within the uncertainties of the experiment we did not resolve any statistically significant heating mechanism. This constrains the collision-related cooling time constant to levels $>5\cdot10^5\,$s, and gives, combined with the thermodynamic treatment reported in Ref.~\onlinecite{kretzschmar2008calculating}, a pressure estimate of $p<10^{-15}\,$mbar. More accurate pressure estimates will become possible by antiproton lifetime measurements, which will be one of the primary objectives of the next antiproton run.

\section{Conclusions}

We presented the development and detailed characterisation of a micrometer thick cryogenic vacuum window based on aluminised Mylar in a wide 900 nm to 2160 nm thickness range which sustains a pressure difference of up to 1300$\,$mbar. 

Various optimisation and endurance tests indicate that the window properties do not degrade after one year of operation at 4~K temperature. Using this window we managed to reach a leak rate smaller than $6.57\pm0.05 \times 10^{-10}$ mbar$\cdot$l/s measured at cryogenic temperature at 1 bar helium pressure, with a foil of only 1760 nm thickness. According to measured degrading properties of the Mylar foil the designed system enables us to transport a low energy antiproton beam provided by AD/ELENA facility at CERN to inside of the apparatus. For 1760 nm foil and 5 kV potential on the trapping electrodes, we expect to be able to trap up to 2.5 \% of the 100 keV antiproton beam. Given all the presented results, we have shown that the developed system meets the requirements for lossless antiproton trapping for at least 12 months.

\section*{Data Availability Statement}

The data that support the findings of this study are available from the corresponding author upon reasonable request.

\begin{acknowledgments}
We acknowledge technical support by CERN, especially the Antiproton Decelerator operation group, CERN's cryolab team and engineering department, and all other CERN groups which provide support to Antiproton Decelerator experiments. 
We acknowledge financial support by RIKEN, the RIKEN SPDR and JRA program, the Max-Planck Society, the European Union (FunI-832848,  STEP-852818), CRC 1227 "DQ-mat"(DFG 274200144), the Cluster of Excellence "Quantum Frontiers" (DFG 390837967), Heinrich Heine University D{\"u}sseldorf, the CERN fellowship program and the CERN Gentner program. This work was supported by the Max-Planck, RIKEN, PTB Center for Time, Constants, and Fundamental Symmetries (C-TCFS).
\end{acknowledgments}

\nocite{*}
\bibliography{2022_degrader_paper_AIP_style_2.bib}

\providecommand{\noopsort}[1]{}\providecommand{\singleletter}[1]{#1}%
\begin{thebibliography}{38}%
\makeatletter
\providecommand \@ifxundefined [1]{%
 \@ifx{#1\undefined}
}%
\providecommand \@ifnum [1]{%
 \ifnum #1\expandafter \@firstoftwo
 \else \expandafter \@secondoftwo
 \fi
}%
\providecommand \@ifx [1]{%
 \ifx #1\expandafter \@firstoftwo
 \else \expandafter \@secondoftwo
 \fi
}%
\providecommand \natexlab [1]{#1}%
\providecommand \enquote  [1]{``#1''}%
\providecommand \bibnamefont  [1]{#1}%
\providecommand \bibfnamefont [1]{#1}%
\providecommand \citenamefont [1]{#1}%
\providecommand \href@noop [0]{\@secondoftwo}%
\providecommand \href [0]{\begingroup \@sanitize@url \@href}%
\providecommand \@href[1]{\@@startlink{#1}\@@href}%
\providecommand \@@href[1]{\endgroup#1\@@endlink}%
\providecommand \@sanitize@url [0]{\catcode `\\12\catcode `\$12\catcode
  `\&12\catcode `\#12\catcode `\^12\catcode `\_12\catcode `\%12\relax}%
\providecommand \@@startlink[1]{}%
\providecommand \@@endlink[0]{}%
\providecommand \url  [0]{\begingroup\@sanitize@url \@url }%
\providecommand \@url [1]{\endgroup\@href {#1}{\urlprefix }}%
\providecommand \urlprefix  [0]{URL }%
\providecommand \Eprint [0]{\href }%
\providecommand \doibase [0]{http://dx.doi.org/}%
\providecommand \selectlanguage [0]{\@gobble}%
\providecommand \bibinfo  [0]{\@secondoftwo}%
\providecommand \bibfield  [0]{\@secondoftwo}%
\providecommand \translation [1]{[#1]}%
\providecommand \BibitemOpen [0]{}%
\providecommand \bibitemStop [0]{}%
\providecommand \bibitemNoStop [0]{.\EOS\space}%
\providecommand \EOS [0]{\spacefactor3000\relax}%
\providecommand \BibitemShut  [1]{\csname bibitem#1\endcsname}%
\let\auto@bib@innerbib\@empty
\bibitem [{\citenamefont {S{\'o}t{\'e}r}\ \emph {et~al.}(2022)\citenamefont
  {S{\'o}t{\'e}r}, \citenamefont {Aghai-Khozani}, \citenamefont {Barna},
  \citenamefont {Dax}, \citenamefont {Venturelli},\ and\ \citenamefont
  {Hori}}]{soter2022high}%
  \BibitemOpen
  \bibfield  {author} {\bibinfo {author} {\bibfnamefont {A.}~\bibnamefont
  {S{\'o}t{\'e}r}}, \bibinfo {author} {\bibfnamefont {H.}~\bibnamefont
  {Aghai-Khozani}}, \bibinfo {author} {\bibfnamefont {D.}~\bibnamefont
  {Barna}}, \bibinfo {author} {\bibfnamefont {A.}~\bibnamefont {Dax}}, \bibinfo
  {author} {\bibfnamefont {L.}~\bibnamefont {Venturelli}}, \ and\ \bibinfo
  {author} {\bibfnamefont {M.}~\bibnamefont {Hori}},\ }\bibfield  {title}
  {\enquote {\bibinfo {title} {High-resolution laser resonances of antiprotonic
  helium in superfluid 4he},}\ }\href@noop {} {\bibfield  {journal} {\bibinfo
  {journal} {Nature}\ }\textbf {\bibinfo {volume} {603}},\ \bibinfo {pages}
  {411--415} (\bibinfo {year} {2022})}\BibitemShut {NoStop}%
\bibitem [{\citenamefont {Tajima}\ and\ \citenamefont
  {Dawson}(1979)}]{tajima1979laser}%
  \BibitemOpen
  \bibfield  {author} {\bibinfo {author} {\bibfnamefont {T.}~\bibnamefont
  {Tajima}}\ and\ \bibinfo {author} {\bibfnamefont {J.~M.}\ \bibnamefont
  {Dawson}},\ }\bibfield  {title} {\enquote {\bibinfo {title} {Laser electron
  accelerator},}\ }\href@noop {} {\bibfield  {journal} {\bibinfo  {journal}
  {Physical Review Letters}\ }\textbf {\bibinfo {volume} {43}},\ \bibinfo
  {pages} {267} (\bibinfo {year} {1979})}\BibitemShut {NoStop}%
\bibitem [{\citenamefont {Chen}\ \emph {et~al.}(1985)\citenamefont {Chen},
  \citenamefont {Dawson}, \citenamefont {Huff},\ and\ \citenamefont
  {Katsouleas}}]{chen1985}%
  \BibitemOpen
  \bibfield  {author} {\bibinfo {author} {\bibfnamefont {P.}~\bibnamefont
  {Chen}}, \bibinfo {author} {\bibfnamefont {J.}~\bibnamefont {Dawson}},
  \bibinfo {author} {\bibfnamefont {R.~W.}\ \bibnamefont {Huff}}, \ and\
  \bibinfo {author} {\bibfnamefont {T.}~\bibnamefont {Katsouleas}},\ }\bibfield
   {title} {\enquote {\bibinfo {title} {Acceleration of electrons by the
  interaction of a bunched electron beam with a plasma [phys. rev. lett. 54,
  693 (1985)]},}\ }\href@noop {} {\bibfield  {journal} {\bibinfo  {journal}
  {Physical Review Letters}\ }\textbf {\bibinfo {volume} {55}},\ \bibinfo
  {pages} {1537} (\bibinfo {year} {1985})}\BibitemShut {NoStop}%
\bibitem [{\citenamefont {Aschikhin}\ \emph {et~al.}(2016)\citenamefont
  {Aschikhin}, \citenamefont {Behrens}, \citenamefont {Bohlen}, \citenamefont
  {Dale}, \citenamefont {Delbos}, \citenamefont {{di Lucchio}}, \citenamefont
  {Elsen}, \citenamefont {Erbe}, \citenamefont {Felber}, \citenamefont
  {Foster}, \citenamefont {Goldberg}, \citenamefont {Grebenyuk}, \citenamefont
  {Gruse}, \citenamefont {Hidding}, \citenamefont {Hu}, \citenamefont
  {Karstensen}, \citenamefont {Knetsch}, \citenamefont {Kononenko},
  \citenamefont {Libov}, \citenamefont {Ludwig}, \citenamefont {Maier},
  \citenamefont {{Martinez de la Ossa}}, \citenamefont {Mehrling},
  \citenamefont {Palmer}, \citenamefont {Pannek}, \citenamefont {Schaper},
  \citenamefont {Schlarb}, \citenamefont {Schmidt}, \citenamefont {Schreiber},
  \citenamefont {Schwinkendorf}, \citenamefont {Steel}, \citenamefont
  {Streeter}, \citenamefont {Tauscher}, \citenamefont {Wacker}, \citenamefont
  {Weichert}, \citenamefont {Wunderlich}, \citenamefont {Zemella},\ and\
  \citenamefont {Osterhoff}}]{ASCHIKHIN2016175}%
  \BibitemOpen
  \bibfield  {author} {\bibinfo {author} {\bibfnamefont {A.}~\bibnamefont
  {Aschikhin}}, \bibinfo {author} {\bibfnamefont {C.}~\bibnamefont {Behrens}},
  \bibinfo {author} {\bibfnamefont {S.}~\bibnamefont {Bohlen}}, \bibinfo
  {author} {\bibfnamefont {J.}~\bibnamefont {Dale}}, \bibinfo {author}
  {\bibfnamefont {N.}~\bibnamefont {Delbos}}, \bibinfo {author} {\bibfnamefont
  {L.}~\bibnamefont {{di Lucchio}}}, \bibinfo {author} {\bibfnamefont
  {E.}~\bibnamefont {Elsen}}, \bibinfo {author} {\bibfnamefont {J.-H.}\
  \bibnamefont {Erbe}}, \bibinfo {author} {\bibfnamefont {M.}~\bibnamefont
  {Felber}}, \bibinfo {author} {\bibfnamefont {B.}~\bibnamefont {Foster}},
  \bibinfo {author} {\bibfnamefont {L.}~\bibnamefont {Goldberg}}, \bibinfo
  {author} {\bibfnamefont {J.}~\bibnamefont {Grebenyuk}}, \bibinfo {author}
  {\bibfnamefont {J.-N.}\ \bibnamefont {Gruse}}, \bibinfo {author}
  {\bibfnamefont {B.}~\bibnamefont {Hidding}}, \bibinfo {author} {\bibfnamefont
  {Z.}~\bibnamefont {Hu}}, \bibinfo {author} {\bibfnamefont {S.}~\bibnamefont
  {Karstensen}}, \bibinfo {author} {\bibfnamefont {A.}~\bibnamefont {Knetsch}},
  \bibinfo {author} {\bibfnamefont {O.}~\bibnamefont {Kononenko}}, \bibinfo
  {author} {\bibfnamefont {V.}~\bibnamefont {Libov}}, \bibinfo {author}
  {\bibfnamefont {K.}~\bibnamefont {Ludwig}}, \bibinfo {author} {\bibfnamefont
  {A.}~\bibnamefont {Maier}}, \bibinfo {author} {\bibfnamefont
  {A.}~\bibnamefont {{Martinez de la Ossa}}}, \bibinfo {author} {\bibfnamefont
  {T.}~\bibnamefont {Mehrling}}, \bibinfo {author} {\bibfnamefont
  {C.}~\bibnamefont {Palmer}}, \bibinfo {author} {\bibfnamefont
  {F.}~\bibnamefont {Pannek}}, \bibinfo {author} {\bibfnamefont
  {L.}~\bibnamefont {Schaper}}, \bibinfo {author} {\bibfnamefont
  {H.}~\bibnamefont {Schlarb}}, \bibinfo {author} {\bibfnamefont
  {B.}~\bibnamefont {Schmidt}}, \bibinfo {author} {\bibfnamefont
  {S.}~\bibnamefont {Schreiber}}, \bibinfo {author} {\bibfnamefont {J.-P.}\
  \bibnamefont {Schwinkendorf}}, \bibinfo {author} {\bibfnamefont
  {H.}~\bibnamefont {Steel}}, \bibinfo {author} {\bibfnamefont
  {M.}~\bibnamefont {Streeter}}, \bibinfo {author} {\bibfnamefont
  {G.}~\bibnamefont {Tauscher}}, \bibinfo {author} {\bibfnamefont
  {V.}~\bibnamefont {Wacker}}, \bibinfo {author} {\bibfnamefont
  {S.}~\bibnamefont {Weichert}}, \bibinfo {author} {\bibfnamefont
  {S.}~\bibnamefont {Wunderlich}}, \bibinfo {author} {\bibfnamefont
  {J.}~\bibnamefont {Zemella}}, \ and\ \bibinfo {author} {\bibfnamefont
  {J.}~\bibnamefont {Osterhoff}},\ }\bibfield  {title} {\enquote {\bibinfo
  {title} {The flashforward facility at desy},}\ }\href {\doibase
  https://doi.org/10.1016/j.nima.2015.10.005} {\bibfield  {journal} {\bibinfo
  {journal} {Nuclear Instruments and Methods in Physics Research Section A:
  Accelerators, Spectrometers, Detectors and Associated Equipment}\ }\textbf
  {\bibinfo {volume} {806}},\ \bibinfo {pages} {175--183} (\bibinfo {year}
  {2016})}\BibitemShut {NoStop}%
\bibitem [{\citenamefont {Smorra}\ \emph
  {et~al.}(2015{\natexlab{a}})\citenamefont {Smorra}, \citenamefont {Blaum},
  \citenamefont {Bojtar}, \citenamefont {Borchert}, \citenamefont {Franke},
  \citenamefont {Higuchi}, \citenamefont {Leefer}, \citenamefont {Nagahama},
  \citenamefont {Matsuda}, \citenamefont {Mooser}, \citenamefont {Niemann},
  \citenamefont {Ospelkaus}, \citenamefont {Quint}, \citenamefont {Schneider},
  \citenamefont {Sellner}, \citenamefont {Tanaka}, \citenamefont {Gorp},
  \citenamefont {Walz}, \citenamefont {Yamazaki},\ and\ \citenamefont
  {Ulmer}}]{Smorra2015BASEExperiment}%
  \BibitemOpen
  \bibfield  {author} {\bibinfo {author} {\bibfnamefont {C.}~\bibnamefont
  {Smorra}}, \bibinfo {author} {\bibfnamefont {K.}~\bibnamefont {Blaum}},
  \bibinfo {author} {\bibfnamefont {L.}~\bibnamefont {Bojtar}}, \bibinfo
  {author} {\bibfnamefont {M.}~\bibnamefont {Borchert}}, \bibinfo {author}
  {\bibfnamefont {K.}~\bibnamefont {Franke}}, \bibinfo {author} {\bibfnamefont
  {T.}~\bibnamefont {Higuchi}}, \bibinfo {author} {\bibfnamefont
  {N.}~\bibnamefont {Leefer}}, \bibinfo {author} {\bibfnamefont
  {H.}~\bibnamefont {Nagahama}}, \bibinfo {author} {\bibfnamefont
  {Y.}~\bibnamefont {Matsuda}}, \bibinfo {author} {\bibfnamefont
  {A.}~\bibnamefont {Mooser}}, \bibinfo {author} {\bibfnamefont
  {M.}~\bibnamefont {Niemann}}, \bibinfo {author} {\bibfnamefont
  {C.}~\bibnamefont {Ospelkaus}}, \bibinfo {author} {\bibfnamefont
  {W.}~\bibnamefont {Quint}}, \bibinfo {author} {\bibfnamefont
  {G.}~\bibnamefont {Schneider}}, \bibinfo {author} {\bibfnamefont
  {S.}~\bibnamefont {Sellner}}, \bibinfo {author} {\bibfnamefont
  {T.}~\bibnamefont {Tanaka}}, \bibinfo {author} {\bibfnamefont {S.~V.}\
  \bibnamefont {Gorp}}, \bibinfo {author} {\bibfnamefont {J.}~\bibnamefont
  {Walz}}, \bibinfo {author} {\bibfnamefont {Y.}~\bibnamefont {Yamazaki}}, \
  and\ \bibinfo {author} {\bibfnamefont {S.}~\bibnamefont {Ulmer}},\ }\bibfield
   {title} {\enquote {\bibinfo {title} {{BASE – The Baryon Antibaryon
  Symmetry Experiment}},}\ }\href {\doibase 10.1140/EPJST/E2015-02607-4}
  {\bibfield  {journal} {\bibinfo  {journal} {The European Physical Journal
  Special Topics}\ }\textbf {\bibinfo {volume} {224}},\ \bibinfo {pages}
  {3055--3108} (\bibinfo {year} {2015}{\natexlab{a}})}\BibitemShut {NoStop}%
\bibitem [{\citenamefont {Smorra}\ \emph {et~al.}(2023)\citenamefont {Smorra},
  \citenamefont {Abbass}, \citenamefont {Bohman}, \citenamefont {Dutheil},
  \citenamefont {Hobl}, \citenamefont {Popper}, \citenamefont {Arndt},
  \citenamefont {Bauer}, \citenamefont {Devlin}, \citenamefont {Erlewein} \emph
  {et~al.}}]{smorra2023base}%
  \BibitemOpen
  \bibfield  {author} {\bibinfo {author} {\bibfnamefont {C.}~\bibnamefont
  {Smorra}}, \bibinfo {author} {\bibfnamefont {F.}~\bibnamefont {Abbass}},
  \bibinfo {author} {\bibfnamefont {M.}~\bibnamefont {Bohman}}, \bibinfo
  {author} {\bibfnamefont {Y.}~\bibnamefont {Dutheil}}, \bibinfo {author}
  {\bibfnamefont {A.}~\bibnamefont {Hobl}}, \bibinfo {author} {\bibfnamefont
  {D.}~\bibnamefont {Popper}}, \bibinfo {author} {\bibfnamefont
  {B.}~\bibnamefont {Arndt}}, \bibinfo {author} {\bibfnamefont
  {B.}~\bibnamefont {Bauer}}, \bibinfo {author} {\bibfnamefont
  {J.}~\bibnamefont {Devlin}}, \bibinfo {author} {\bibfnamefont
  {S.}~\bibnamefont {Erlewein}},  \emph {et~al.},\ }\bibfield  {title}
  {\enquote {\bibinfo {title} {Base-step: A transportable antiproton reservoir
  for fundamental interaction studies},}\ }\href@noop {} {\bibfield  {journal}
  {\bibinfo  {journal} {arXiv preprint arXiv:2304.09555}\ } (\bibinfo {year}
  {2023})}\BibitemShut {NoStop}%
\bibitem [{\citenamefont {Gabrielse}\ \emph {et~al.}(2002)\citenamefont
  {Gabrielse}, \citenamefont {Bowden}, \citenamefont {Oxley}, \citenamefont
  {Speck}, \citenamefont {Storry}, \citenamefont {Tan}, \citenamefont
  {Wessels}, \citenamefont {Grzonka}, \citenamefont {Oelert}, \citenamefont
  {Schepers} \emph {et~al.}}]{gabrielse2002stacking}%
  \BibitemOpen
  \bibfield  {author} {\bibinfo {author} {\bibfnamefont {G.}~\bibnamefont
  {Gabrielse}}, \bibinfo {author} {\bibfnamefont {N.}~\bibnamefont {Bowden}},
  \bibinfo {author} {\bibfnamefont {P.}~\bibnamefont {Oxley}}, \bibinfo
  {author} {\bibfnamefont {A.}~\bibnamefont {Speck}}, \bibinfo {author}
  {\bibfnamefont {C.}~\bibnamefont {Storry}}, \bibinfo {author} {\bibfnamefont
  {J.}~\bibnamefont {Tan}}, \bibinfo {author} {\bibfnamefont {M.}~\bibnamefont
  {Wessels}}, \bibinfo {author} {\bibfnamefont {D.}~\bibnamefont {Grzonka}},
  \bibinfo {author} {\bibfnamefont {W.}~\bibnamefont {Oelert}}, \bibinfo
  {author} {\bibfnamefont {G.}~\bibnamefont {Schepers}},  \emph {et~al.},\
  }\bibfield  {title} {\enquote {\bibinfo {title} {Stacking of cold
  antiprotons},}\ }\href@noop {} {\bibfield  {journal} {\bibinfo  {journal}
  {Physics Letters B}\ }\textbf {\bibinfo {volume} {548}},\ \bibinfo {pages}
  {140--145} (\bibinfo {year} {2002})}\BibitemShut {NoStop}%
\bibitem [{\citenamefont {Kuroda}\ \emph {et~al.}(2005)\citenamefont {Kuroda},
  \citenamefont {Torii}, \citenamefont {Franzen}, \citenamefont {Wang},
  \citenamefont {Yoneda}, \citenamefont {Inoue}, \citenamefont {Hori},
  \citenamefont {Juh{\'a}sz}, \citenamefont {Horv{\'a}th}, \citenamefont
  {Higaki} \emph {et~al.}}]{kuroda2005confinement}%
  \BibitemOpen
  \bibfield  {author} {\bibinfo {author} {\bibfnamefont {N.}~\bibnamefont
  {Kuroda}}, \bibinfo {author} {\bibfnamefont {H.~A.}\ \bibnamefont {Torii}},
  \bibinfo {author} {\bibfnamefont {K.~Y.}\ \bibnamefont {Franzen}}, \bibinfo
  {author} {\bibfnamefont {Z.}~\bibnamefont {Wang}}, \bibinfo {author}
  {\bibfnamefont {S.}~\bibnamefont {Yoneda}}, \bibinfo {author} {\bibfnamefont
  {M.}~\bibnamefont {Inoue}}, \bibinfo {author} {\bibfnamefont
  {M.}~\bibnamefont {Hori}}, \bibinfo {author} {\bibfnamefont {B.}~\bibnamefont
  {Juh{\'a}sz}}, \bibinfo {author} {\bibfnamefont {D.}~\bibnamefont
  {Horv{\'a}th}}, \bibinfo {author} {\bibfnamefont {H.}~\bibnamefont {Higaki}},
   \emph {et~al.},\ }\bibfield  {title} {\enquote {\bibinfo {title}
  {Confinement of a large number of antiprotons and production of an ultraslow
  antiproton beam},}\ }\href@noop {} {\bibfield  {journal} {\bibinfo  {journal}
  {Physical review letters}\ }\textbf {\bibinfo {volume} {94}},\ \bibinfo
  {pages} {023401} (\bibinfo {year} {2005})}\BibitemShut {NoStop}%
\bibitem [{\citenamefont {Amole}\ \emph {et~al.}(2014)\citenamefont {Amole},
  \citenamefont {Andresen}, \citenamefont {Ashkezari}, \citenamefont
  {Baquero-Ruiz}, \citenamefont {Bertsche}, \citenamefont {Bowe}, \citenamefont
  {Butler}, \citenamefont {Capra}, \citenamefont {Carpenter}, \citenamefont
  {Cesar} \emph {et~al.}}]{amole2014alpha}%
  \BibitemOpen
  \bibfield  {author} {\bibinfo {author} {\bibfnamefont {C.}~\bibnamefont
  {Amole}}, \bibinfo {author} {\bibfnamefont {G.}~\bibnamefont {Andresen}},
  \bibinfo {author} {\bibfnamefont {M.}~\bibnamefont {Ashkezari}}, \bibinfo
  {author} {\bibfnamefont {M.}~\bibnamefont {Baquero-Ruiz}}, \bibinfo {author}
  {\bibfnamefont {W.}~\bibnamefont {Bertsche}}, \bibinfo {author}
  {\bibfnamefont {P.}~\bibnamefont {Bowe}}, \bibinfo {author} {\bibfnamefont
  {E.}~\bibnamefont {Butler}}, \bibinfo {author} {\bibfnamefont
  {A.}~\bibnamefont {Capra}}, \bibinfo {author} {\bibfnamefont
  {P.}~\bibnamefont {Carpenter}}, \bibinfo {author} {\bibfnamefont
  {C.}~\bibnamefont {Cesar}},  \emph {et~al.},\ }\bibfield  {title} {\enquote
  {\bibinfo {title} {The alpha antihydrogen trapping apparatus},}\ }\href@noop
  {} {\bibfield  {journal} {\bibinfo  {journal} {Nuclear Instruments and
  Methods in Physics Research Section A: Accelerators, Spectrometers, Detectors
  and Associated Equipment}\ }\textbf {\bibinfo {volume} {735}},\ \bibinfo
  {pages} {319--340} (\bibinfo {year} {2014})}\BibitemShut {NoStop}%
\bibitem [{\citenamefont {Gutierrez}\ \emph {et~al.}(2015)\citenamefont
  {Gutierrez}, \citenamefont {Ashkezari}, \citenamefont {Baquero-Ruiz} \emph
  {et~al.}}]{amole2015alpha}%
  \BibitemOpen
  \bibfield  {author} {\bibinfo {author} {\bibfnamefont {A.}~\bibnamefont
  {Gutierrez}}, \bibinfo {author} {\bibfnamefont {M.}~\bibnamefont
  {Ashkezari}}, \bibinfo {author} {\bibfnamefont {M.}~\bibnamefont
  {Baquero-Ruiz}},  \emph {et~al.},\ }\bibfield  {title} {\enquote {\bibinfo
  {title} {Antiproton cloud compression in the alpha apparatus at cern.}}\
  }\href {\doibase 10.1007/s10751-015-1202-4} {\bibfield  {journal} {\bibinfo
  {journal} {Hyperfine Interact}\ }\textbf {\bibinfo {volume} {235}},\ \bibinfo
  {pages} {21--28} (\bibinfo {year} {2015})}\BibitemShut {NoStop}%
\bibitem [{\citenamefont {Scampoli}\ and\ \citenamefont
  {Storey}(2014)}]{scampoli2014aegis}%
  \BibitemOpen
  \bibfield  {author} {\bibinfo {author} {\bibfnamefont {P.}~\bibnamefont
  {Scampoli}}\ and\ \bibinfo {author} {\bibfnamefont {J.}~\bibnamefont
  {Storey}},\ }\bibfield  {title} {\enquote {\bibinfo {title} {The aegis
  experiment at cern for the measurement of antihydrogen gravity
  acceleration},}\ }\href@noop {} {\bibfield  {journal} {\bibinfo  {journal}
  {Modern Physics Letters A}\ }\textbf {\bibinfo {volume} {29}},\ \bibinfo
  {pages} {1430017} (\bibinfo {year} {2014})}\BibitemShut {NoStop}%
\bibitem [{\citenamefont {Kellerbauer}\ \emph {et~al.}(2001)\citenamefont
  {Kellerbauer}, \citenamefont {Kim}, \citenamefont {Moore},\ and\
  \citenamefont {Varfalvy}}]{kellerbauer2001buffer}%
  \BibitemOpen
  \bibfield  {author} {\bibinfo {author} {\bibfnamefont {A.}~\bibnamefont
  {Kellerbauer}}, \bibinfo {author} {\bibfnamefont {T.}~\bibnamefont {Kim}},
  \bibinfo {author} {\bibfnamefont {R.}~\bibnamefont {Moore}}, \ and\ \bibinfo
  {author} {\bibfnamefont {P.}~\bibnamefont {Varfalvy}},\ }\bibfield  {title}
  {\enquote {\bibinfo {title} {Buffer gas cooling of ion beams},}\ }\href@noop
  {} {\bibfield  {journal} {\bibinfo  {journal} {Nuclear Instruments and
  Methods in Physics Research Section A: Accelerators, Spectrometers, Detectors
  and Associated Equipment}\ }\textbf {\bibinfo {volume} {469}},\ \bibinfo
  {pages} {276--285} (\bibinfo {year} {2001})}\BibitemShut {NoStop}%
\bibitem [{\citenamefont {Droese}\ \emph {et~al.}(2014)\citenamefont {Droese},
  \citenamefont {Eliseev}, \citenamefont {Blaum}, \citenamefont {Block},
  \citenamefont {Herfurth}, \citenamefont {Laatiaoui}, \citenamefont
  {Lautenschl{\"a}ger}, \citenamefont {Ramirez}, \citenamefont {Schweikhard},
  \citenamefont {Simon} \emph {et~al.}}]{droese2014cryogenic}%
  \BibitemOpen
  \bibfield  {author} {\bibinfo {author} {\bibfnamefont {C.}~\bibnamefont
  {Droese}}, \bibinfo {author} {\bibfnamefont {S.}~\bibnamefont {Eliseev}},
  \bibinfo {author} {\bibfnamefont {K.}~\bibnamefont {Blaum}}, \bibinfo
  {author} {\bibfnamefont {M.}~\bibnamefont {Block}}, \bibinfo {author}
  {\bibfnamefont {F.}~\bibnamefont {Herfurth}}, \bibinfo {author}
  {\bibfnamefont {M.}~\bibnamefont {Laatiaoui}}, \bibinfo {author}
  {\bibfnamefont {F.}~\bibnamefont {Lautenschl{\"a}ger}}, \bibinfo {author}
  {\bibfnamefont {E.~M.}\ \bibnamefont {Ramirez}}, \bibinfo {author}
  {\bibfnamefont {L.}~\bibnamefont {Schweikhard}}, \bibinfo {author}
  {\bibfnamefont {V.}~\bibnamefont {Simon}},  \emph {et~al.},\ }\bibfield
  {title} {\enquote {\bibinfo {title} {The cryogenic gas stopping cell of
  shiptrap},}\ }\href@noop {} {\bibfield  {journal} {\bibinfo  {journal}
  {Nuclear Instruments and Methods in Physics Research Section B: Beam
  Interactions with Materials and Atoms}\ }\textbf {\bibinfo {volume} {338}},\
  \bibinfo {pages} {126--138} (\bibinfo {year} {2014})}\BibitemShut {NoStop}%
\bibitem [{\citenamefont {Ulbricht}(2006)}]{ULBRICHT20062217}%
  \BibitemOpen
  \bibfield  {author} {\bibinfo {author} {\bibfnamefont {M.}~\bibnamefont
  {Ulbricht}},\ }\bibfield  {title} {\enquote {\bibinfo {title} {Advanced
  functional polymer membranes},}\ }\href {\doibase
  https://doi.org/10.1016/j.polymer.2006.01.084} {\bibfield  {journal}
  {\bibinfo  {journal} {Polymer}\ }\textbf {\bibinfo {volume} {47}},\ \bibinfo
  {pages} {2217--2262} (\bibinfo {year} {2006})},\ \bibinfo {note} {single
  Chain Polymers}\BibitemShut {NoStop}%
\bibitem [{\citenamefont {Crowley}\ and\ \citenamefont
  {Pizziconi}(2005)}]{crowley2005isolation}%
  \BibitemOpen
  \bibfield  {author} {\bibinfo {author} {\bibfnamefont {T.~A.}\ \bibnamefont
  {Crowley}}\ and\ \bibinfo {author} {\bibfnamefont {V.}~\bibnamefont
  {Pizziconi}},\ }\bibfield  {title} {\enquote {\bibinfo {title} {Isolation of
  plasma from whole blood using planar microfilters for lab-on-a-chip
  applications},}\ }\href@noop {} {\bibfield  {journal} {\bibinfo  {journal}
  {Lab on a Chip}\ }\textbf {\bibinfo {volume} {5}},\ \bibinfo {pages}
  {922--929} (\bibinfo {year} {2005})}\BibitemShut {NoStop}%
\bibitem [{\citenamefont {Ehrenhofer}\ \emph {et~al.}(2016)\citenamefont
  {Ehrenhofer}, \citenamefont {Bingel}, \citenamefont {Paschew}, \citenamefont
  {Tietze}, \citenamefont {Schr{\"o}der}, \citenamefont {Richter},\ and\
  \citenamefont {Wallmersperger}}]{ehrenhofer2016permeation}%
  \BibitemOpen
  \bibfield  {author} {\bibinfo {author} {\bibfnamefont {A.}~\bibnamefont
  {Ehrenhofer}}, \bibinfo {author} {\bibfnamefont {G.}~\bibnamefont {Bingel}},
  \bibinfo {author} {\bibfnamefont {G.}~\bibnamefont {Paschew}}, \bibinfo
  {author} {\bibfnamefont {M.}~\bibnamefont {Tietze}}, \bibinfo {author}
  {\bibfnamefont {R.}~\bibnamefont {Schr{\"o}der}}, \bibinfo {author}
  {\bibfnamefont {A.}~\bibnamefont {Richter}}, \ and\ \bibinfo {author}
  {\bibfnamefont {T.}~\bibnamefont {Wallmersperger}},\ }\bibfield  {title}
  {\enquote {\bibinfo {title} {Permeation control in hydrogel-layered patterned
  pet membranes with defined switchable pore geometry--experiments and
  numerical simulation},}\ }\href@noop {} {\bibfield  {journal} {\bibinfo
  {journal} {Sensors and Actuators B: Chemical}\ }\textbf {\bibinfo {volume}
  {232}},\ \bibinfo {pages} {499--505} (\bibinfo {year} {2016})}\BibitemShut
  {NoStop}%
\bibitem [{\citenamefont {Maury}(1997)}]{maury1997antiproton}%
  \BibitemOpen
  \bibfield  {author} {\bibinfo {author} {\bibfnamefont {S.}~\bibnamefont
  {Maury}},\ }\bibfield  {title} {\enquote {\bibinfo {title} {The antiproton
  decelerator: Ad},}\ }\href@noop {} {\bibfield  {journal} {\bibinfo  {journal}
  {Hyperfine Interactions}\ }\textbf {\bibinfo {volume} {109}},\ \bibinfo
  {pages} {43--52} (\bibinfo {year} {1997})}\BibitemShut {NoStop}%
\bibitem [{\citenamefont {Smorra}\ \emph {et~al.}(2017)\citenamefont {Smorra},
  \citenamefont {Sellner}, \citenamefont {Borchert}, \citenamefont
  {Harrington}, \citenamefont {Higuchi}, \citenamefont {Nagahama},
  \citenamefont {Tanaka}, \citenamefont {Mooser}, \citenamefont {Schneider},
  \citenamefont {Bohman} \emph {et~al.}}]{BASE_antip_mag_moment}%
  \BibitemOpen
  \bibfield  {author} {\bibinfo {author} {\bibfnamefont {C.}~\bibnamefont
  {Smorra}}, \bibinfo {author} {\bibfnamefont {S.}~\bibnamefont {Sellner}},
  \bibinfo {author} {\bibfnamefont {M.}~\bibnamefont {Borchert}}, \bibinfo
  {author} {\bibfnamefont {J.}~\bibnamefont {Harrington}}, \bibinfo {author}
  {\bibfnamefont {T.}~\bibnamefont {Higuchi}}, \bibinfo {author} {\bibfnamefont
  {H.}~\bibnamefont {Nagahama}}, \bibinfo {author} {\bibfnamefont
  {T.}~\bibnamefont {Tanaka}}, \bibinfo {author} {\bibfnamefont
  {A.}~\bibnamefont {Mooser}}, \bibinfo {author} {\bibfnamefont
  {G.}~\bibnamefont {Schneider}}, \bibinfo {author} {\bibfnamefont
  {M.}~\bibnamefont {Bohman}},  \emph {et~al.},\ }\bibfield  {title} {\enquote
  {\bibinfo {title} {A parts-per-billion measurement of the antiproton magnetic
  moment},}\ }\href@noop {} {\bibfield  {journal} {\bibinfo  {journal}
  {Nature}\ }\textbf {\bibinfo {volume} {550}},\ \bibinfo {pages} {371--374}
  (\bibinfo {year} {2017})}\BibitemShut {NoStop}%
\bibitem [{\citenamefont {Borchert}\ \emph {et~al.}(2022)\citenamefont
  {Borchert}, \citenamefont {Devlin}, \citenamefont {Erlewein}, \citenamefont
  {Fleck}, \citenamefont {Harrington}, \citenamefont {Higuchi}, \citenamefont
  {Latacz}, \citenamefont {Voelksen}, \citenamefont {Wursten}, \citenamefont
  {Abbass} \emph {et~al.}}]{borchert202216}%
  \BibitemOpen
  \bibfield  {author} {\bibinfo {author} {\bibfnamefont {M.}~\bibnamefont
  {Borchert}}, \bibinfo {author} {\bibfnamefont {J.}~\bibnamefont {Devlin}},
  \bibinfo {author} {\bibfnamefont {S.}~\bibnamefont {Erlewein}}, \bibinfo
  {author} {\bibfnamefont {M.}~\bibnamefont {Fleck}}, \bibinfo {author}
  {\bibfnamefont {J.}~\bibnamefont {Harrington}}, \bibinfo {author}
  {\bibfnamefont {T.}~\bibnamefont {Higuchi}}, \bibinfo {author} {\bibfnamefont
  {B.}~\bibnamefont {Latacz}}, \bibinfo {author} {\bibfnamefont
  {F.}~\bibnamefont {Voelksen}}, \bibinfo {author} {\bibfnamefont
  {E.}~\bibnamefont {Wursten}}, \bibinfo {author} {\bibfnamefont
  {F.}~\bibnamefont {Abbass}},  \emph {et~al.},\ }\bibfield  {title} {\enquote
  {\bibinfo {title} {A 16-parts-per-trillion measurement of the
  antiproton-to-proton charge--mass ratio},}\ }\href@noop {} {\bibfield
  {journal} {\bibinfo  {journal} {Nature}\ }\textbf {\bibinfo {volume} {601}},\
  \bibinfo {pages} {53--57} (\bibinfo {year} {2022})}\BibitemShut {NoStop}%
\bibitem [{\citenamefont {Sellner}\ \emph {et~al.}(2017)\citenamefont
  {Sellner}, \citenamefont {Besirli}, \citenamefont {Bohman}, \citenamefont
  {Borchert}, \citenamefont {Harrington}, \citenamefont {Higuchi},
  \citenamefont {Mooser}, \citenamefont {Nagahama}, \citenamefont {Schneider},
  \citenamefont {Smorra} \emph {et~al.}}]{sellner2017improved}%
  \BibitemOpen
  \bibfield  {author} {\bibinfo {author} {\bibfnamefont {S.}~\bibnamefont
  {Sellner}}, \bibinfo {author} {\bibfnamefont {M.}~\bibnamefont {Besirli}},
  \bibinfo {author} {\bibfnamefont {M.}~\bibnamefont {Bohman}}, \bibinfo
  {author} {\bibfnamefont {M.}~\bibnamefont {Borchert}}, \bibinfo {author}
  {\bibfnamefont {J.}~\bibnamefont {Harrington}}, \bibinfo {author}
  {\bibfnamefont {T.}~\bibnamefont {Higuchi}}, \bibinfo {author} {\bibfnamefont
  {A.}~\bibnamefont {Mooser}}, \bibinfo {author} {\bibfnamefont
  {H.}~\bibnamefont {Nagahama}}, \bibinfo {author} {\bibfnamefont
  {G.}~\bibnamefont {Schneider}}, \bibinfo {author} {\bibfnamefont
  {C.}~\bibnamefont {Smorra}},  \emph {et~al.},\ }\bibfield  {title} {\enquote
  {\bibinfo {title} {Improved limit on the directly measured antiproton
  lifetime},}\ }\href@noop {} {\bibfield  {journal} {\bibinfo  {journal} {New
  Journal of Physics}\ }\textbf {\bibinfo {volume} {19}},\ \bibinfo {pages}
  {083023} (\bibinfo {year} {2017})}\BibitemShut {NoStop}%
\bibitem [{\citenamefont {Fei}(1990)}]{Fei}%
  \BibitemOpen
  \bibfield  {author} {\bibinfo {author} {\bibfnamefont {X.}~\bibnamefont
  {Fei}},\ }\emph {\bibinfo {title} {Trapping low energy antiprotons in an ion
  trap}},\ \href@noop {} {Ph.D. thesis},\ \bibinfo  {school} {Harvard
  University, Department of Physics} (\bibinfo {year} {1990})\BibitemShut
  {NoStop}%
\bibitem [{\citenamefont {Kuchler}\ \emph {et~al.}(2014)\citenamefont
  {Kuchler}, \citenamefont {Paoluzzi}, \citenamefont {Hori}, \citenamefont
  {Eriksson}, \citenamefont {Pedersen}, \citenamefont {Vanbavinckhove},
  \citenamefont {Puccio}, \citenamefont {Maury}, \citenamefont {Fedemann},
  \citenamefont {Harasimowicz} \emph {et~al.}}]{kuchler2014extra}%
  \BibitemOpen
  \bibfield  {author} {\bibinfo {author} {\bibfnamefont {D.}~\bibnamefont
  {Kuchler}}, \bibinfo {author} {\bibfnamefont {M.}~\bibnamefont {Paoluzzi}},
  \bibinfo {author} {\bibfnamefont {M.}~\bibnamefont {Hori}}, \bibinfo {author}
  {\bibfnamefont {T.}~\bibnamefont {Eriksson}}, \bibinfo {author}
  {\bibfnamefont {F.}~\bibnamefont {Pedersen}}, \bibinfo {author}
  {\bibfnamefont {G.}~\bibnamefont {Vanbavinckhove}}, \bibinfo {author}
  {\bibfnamefont {B.}~\bibnamefont {Puccio}}, \bibinfo {author} {\bibfnamefont
  {S.}~\bibnamefont {Maury}}, \bibinfo {author} {\bibfnamefont
  {S.}~\bibnamefont {Fedemann}}, \bibinfo {author} {\bibfnamefont
  {J.}~\bibnamefont {Harasimowicz}},  \emph {et~al.},\ }\href@noop {} {\enquote
  {\bibinfo {title} {Extra low energy antiproton (elena) ring and its transfer
  lines: Design report},}\ }\bibinfo {type} {Tech. Rep.}\ (\bibinfo
  {institution} {CERN},\ \bibinfo {year} {2014})\BibitemShut {NoStop}%
\bibitem [{\citenamefont {Engel}\ \emph {et~al.}(2020)\citenamefont {Engel},
  \citenamefont {Gross}, \citenamefont {Koss}, \citenamefont {Lishilin},
  \citenamefont {Loisch}, \citenamefont {Philipp}, \citenamefont {Richter},\
  and\ \citenamefont {Stephan}}]{engel2020polymer}%
  \BibitemOpen
  \bibfield  {author} {\bibinfo {author} {\bibfnamefont {J.}~\bibnamefont
  {Engel}}, \bibinfo {author} {\bibfnamefont {M.}~\bibnamefont {Gross}},
  \bibinfo {author} {\bibfnamefont {G.}~\bibnamefont {Koss}}, \bibinfo {author}
  {\bibfnamefont {O.}~\bibnamefont {Lishilin}}, \bibinfo {author}
  {\bibfnamefont {G.}~\bibnamefont {Loisch}}, \bibinfo {author} {\bibfnamefont
  {S.}~\bibnamefont {Philipp}}, \bibinfo {author} {\bibfnamefont
  {D.}~\bibnamefont {Richter}}, \ and\ \bibinfo {author} {\bibfnamefont
  {F.}~\bibnamefont {Stephan}},\ }\bibfield  {title} {\enquote {\bibinfo
  {title} {Polymer foil windows for gas--vacuum separation in accelerator
  applications},}\ }\href@noop {} {\bibfield  {journal} {\bibinfo  {journal}
  {AIP Advances}\ }\textbf {\bibinfo {volume} {10}},\ \bibinfo {pages} {025224}
  (\bibinfo {year} {2020})}\BibitemShut {NoStop}%
\bibitem [{\citenamefont {Mapes}, \citenamefont {Hseuh},\ and\ \citenamefont
  {Jiang}(1994)}]{mapes1994permeation}%
  \BibitemOpen
  \bibfield  {author} {\bibinfo {author} {\bibfnamefont {M.}~\bibnamefont
  {Mapes}}, \bibinfo {author} {\bibfnamefont {H.}~\bibnamefont {Hseuh}}, \ and\
  \bibinfo {author} {\bibfnamefont {W.}~\bibnamefont {Jiang}},\ }\bibfield
  {title} {\enquote {\bibinfo {title} {Permeation of argon, carbon dioxide,
  helium, nitrogen, and oxygen through mylar windows},}\ }\href@noop {}
  {\bibfield  {journal} {\bibinfo  {journal} {Journal of Vacuum Science \&
  Technology A: Vacuum, Surfaces, and Films}\ }\textbf {\bibinfo {volume}
  {12}},\ \bibinfo {pages} {1699--1704} (\bibinfo {year} {1994})}\BibitemShut
  {NoStop}%
\bibitem [{\citenamefont {Hassenzahl}\ and\ \citenamefont
  {Gray}(1975)}]{HASSENZAHL1975627}%
  \BibitemOpen
  \bibfield  {author} {\bibinfo {author} {\bibfnamefont {W.}~\bibnamefont
  {Hassenzahl}}\ and\ \bibinfo {author} {\bibfnamefont {W.}~\bibnamefont
  {Gray}},\ }\bibfield  {title} {\enquote {\bibinfo {title} {Thin windows for
  gaseous and liquid targets: An optimization procedure},}\ }\href {\doibase
  https://doi.org/10.1016/0011-2275(75)90093-4} {\bibfield  {journal} {\bibinfo
   {journal} {Cryogenics}\ }\textbf {\bibinfo {volume} {15}},\ \bibinfo {pages}
  {627--638} (\bibinfo {year} {1975})}\BibitemShut {NoStop}%
\bibitem [{Api(7 21)}]{ApiezonN}%
  \BibitemOpen
  \href@noop {} {\enquote {\bibinfo {title} {Apiezon, ultra high and high
  vacuum greases},}\ }\bibinfo {howpublished}
  {\url{https://static.mimaterials.com/apiezon/DocumentLibrary/TechnicalDatasheets/Apiezon_L_M_and_N_Ultra_High_and_High_Vacuum_Greases_Datasheet.pdf}}
  (\bibinfo {year} {Accessed: 2022-07-21})\BibitemShut {NoStop}%
\bibitem [{\citenamefont {Smorra}\ \emph
  {et~al.}(2015{\natexlab{b}})\citenamefont {Smorra}, \citenamefont {Blaum},
  \citenamefont {Bojtar}, \citenamefont {Borchert}, \citenamefont {Franke},
  \citenamefont {Higuchi}, \citenamefont {Leefer}, \citenamefont {Nagahama},
  \citenamefont {Matsuda}, \citenamefont {Mooser} \emph
  {et~al.}}]{smorra2015base}%
  \BibitemOpen
  \bibfield  {author} {\bibinfo {author} {\bibfnamefont {C.}~\bibnamefont
  {Smorra}}, \bibinfo {author} {\bibfnamefont {K.}~\bibnamefont {Blaum}},
  \bibinfo {author} {\bibfnamefont {L.}~\bibnamefont {Bojtar}}, \bibinfo
  {author} {\bibfnamefont {M.}~\bibnamefont {Borchert}}, \bibinfo {author}
  {\bibfnamefont {K.}~\bibnamefont {Franke}}, \bibinfo {author} {\bibfnamefont
  {T.}~\bibnamefont {Higuchi}}, \bibinfo {author} {\bibfnamefont
  {N.}~\bibnamefont {Leefer}}, \bibinfo {author} {\bibfnamefont
  {H.}~\bibnamefont {Nagahama}}, \bibinfo {author} {\bibfnamefont
  {Y.}~\bibnamefont {Matsuda}}, \bibinfo {author} {\bibfnamefont
  {A.}~\bibnamefont {Mooser}},  \emph {et~al.},\ }\bibfield  {title} {\enquote
  {\bibinfo {title} {Base--the baryon antibaryon symmetry experiment},}\
  }\href@noop {} {\bibfield  {journal} {\bibinfo  {journal} {The European
  Physical Journal Special Topics}\ }\textbf {\bibinfo {volume} {224}},\
  \bibinfo {pages} {3055--3108} (\bibinfo {year}
  {2015}{\natexlab{b}})}\BibitemShut {NoStop}%
\bibitem [{\citenamefont {Ru{\v{z}}barsk{\`y}}(2022)}]{ruvzbarskycontactless}%
  \BibitemOpen
  \bibfield  {author} {\bibinfo {author} {\bibfnamefont {J.}~\bibnamefont
  {Ru{\v{z}}barsk{\`y}}},\ }\href@noop {} {\emph {\bibinfo {title} {Contactless
  System for Measurement and Evaluation of Machined Surfaces}}}\ (\bibinfo
  {publisher} {Springer},\ \bibinfo {year} {2022})\BibitemShut {NoStop}%
\bibitem [{\citenamefont {Gruen}(1991)}]{gruen1991process}%
  \BibitemOpen
  \bibfield  {author} {\bibinfo {author} {\bibfnamefont {R.}~\bibnamefont
  {Gruen}},\ }\href@noop {} {\enquote {\bibinfo {title} {Process and apparatus
  for coating conducting pieces using a pulsed glow discharge},}\ } (\bibinfo
  {year} {1991}),\ \bibinfo {note} {uS Patent 5,015,493}\BibitemShut {NoStop}%
\bibitem [{\citenamefont {Klauk}(2006)}]{klauk2006organic}%
  \BibitemOpen
  \bibfield  {author} {\bibinfo {author} {\bibfnamefont {H.}~\bibnamefont
  {Klauk}},\ }\href@noop {} {\emph {\bibinfo {title} {Organic electronics:
  materials, manufacturing, and applications}}}\ (\bibinfo  {publisher} {John
  Wiley \& Sons},\ \bibinfo {year} {2006})\BibitemShut {NoStop}%
\bibitem [{\citenamefont {Fuoco}\ \emph {et~al.}(2018)\citenamefont {Fuoco},
  \citenamefont {Comesa{\~n}a-G{\'a}ndara}, \citenamefont {Longo},
  \citenamefont {Esposito}, \citenamefont {Monteleone}, \citenamefont {Rose},
  \citenamefont {Bezzu}, \citenamefont {Carta}, \citenamefont {McKeown},\ and\
  \citenamefont {Jansen}}]{fuoco2018temperature}%
  \BibitemOpen
  \bibfield  {author} {\bibinfo {author} {\bibfnamefont {A.}~\bibnamefont
  {Fuoco}}, \bibinfo {author} {\bibfnamefont {B.}~\bibnamefont
  {Comesa{\~n}a-G{\'a}ndara}}, \bibinfo {author} {\bibfnamefont
  {M.}~\bibnamefont {Longo}}, \bibinfo {author} {\bibfnamefont
  {E.}~\bibnamefont {Esposito}}, \bibinfo {author} {\bibfnamefont
  {M.}~\bibnamefont {Monteleone}}, \bibinfo {author} {\bibfnamefont
  {I.}~\bibnamefont {Rose}}, \bibinfo {author} {\bibfnamefont {C.~G.}\
  \bibnamefont {Bezzu}}, \bibinfo {author} {\bibfnamefont {M.}~\bibnamefont
  {Carta}}, \bibinfo {author} {\bibfnamefont {N.~B.}\ \bibnamefont {McKeown}},
  \ and\ \bibinfo {author} {\bibfnamefont {J.~C.}\ \bibnamefont {Jansen}},\
  }\bibfield  {title} {\enquote {\bibinfo {title} {Temperature dependence of
  gas permeation and diffusion in triptycene-based ultrapermeable polymers of
  intrinsic microporosity},}\ }\href@noop {} {\bibfield  {journal} {\bibinfo
  {journal} {ACS applied materials \& interfaces}\ }\textbf {\bibinfo {volume}
  {10}},\ \bibinfo {pages} {36475--36482} (\bibinfo {year} {2018})}\BibitemShut
  {NoStop}%
\bibitem [{\citenamefont {M{\o}ller}\ \emph {et~al.}(2002)\citenamefont
  {M{\o}ller}, \citenamefont {Csete}, \citenamefont {Ichioka}, \citenamefont
  {Knudsen}, \citenamefont {Uggerh{\o}j},\ and\ \citenamefont
  {Andersen}}]{moller2002antiproton}%
  \BibitemOpen
  \bibfield  {author} {\bibinfo {author} {\bibfnamefont {S.~P.}\ \bibnamefont
  {M{\o}ller}}, \bibinfo {author} {\bibfnamefont {A.}~\bibnamefont {Csete}},
  \bibinfo {author} {\bibfnamefont {T.}~\bibnamefont {Ichioka}}, \bibinfo
  {author} {\bibfnamefont {H.}~\bibnamefont {Knudsen}}, \bibinfo {author}
  {\bibfnamefont {U.}~\bibnamefont {Uggerh{\o}j}}, \ and\ \bibinfo {author}
  {\bibfnamefont {H.}~\bibnamefont {Andersen}},\ }\bibfield  {title} {\enquote
  {\bibinfo {title} {Antiproton stopping at low energies: Confirmation of
  velocity-proportional stopping power},}\ }\href@noop {} {\bibfield  {journal}
  {\bibinfo  {journal} {Physical Review Letters}\ }\textbf {\bibinfo {volume}
  {88}},\ \bibinfo {pages} {193201} (\bibinfo {year} {2002})}\BibitemShut
  {NoStop}%
\bibitem [{\citenamefont {Ziegler}, \citenamefont {Ziegler},\ and\
  \citenamefont {Biersack}(2010)}]{ziegler2010srim}%
  \BibitemOpen
  \bibfield  {author} {\bibinfo {author} {\bibfnamefont {J.~F.}\ \bibnamefont
  {Ziegler}}, \bibinfo {author} {\bibfnamefont {M.~D.}\ \bibnamefont
  {Ziegler}}, \ and\ \bibinfo {author} {\bibfnamefont {J.~P.}\ \bibnamefont
  {Biersack}},\ }\bibfield  {title} {\enquote {\bibinfo {title} {Srim--the
  stopping and range of ions in matter (2010)},}\ }\href@noop {} {\bibfield
  {journal} {\bibinfo  {journal} {Nuclear Instruments and Methods in Physics
  Research Section B: Beam Interactions with Materials and Atoms}\ }\textbf
  {\bibinfo {volume} {268}},\ \bibinfo {pages} {1818--1823} (\bibinfo {year}
  {2010})}\BibitemShut {NoStop}%
\bibitem [{\citenamefont {Thwaites}(1983)}]{thwaites1983bragg}%
  \BibitemOpen
  \bibfield  {author} {\bibinfo {author} {\bibfnamefont {D.}~\bibnamefont
  {Thwaites}},\ }\bibfield  {title} {\enquote {\bibinfo {title} {Bragg's rule
  of stopping power additivity: A compilation and summary of results},}\
  }\href@noop {} {\bibfield  {journal} {\bibinfo  {journal} {Radiation
  Research}\ }\textbf {\bibinfo {volume} {95}},\ \bibinfo {pages} {495--518}
  (\bibinfo {year} {1983})}\BibitemShut {NoStop}%
\bibitem [{\citenamefont {Nordlund}, \citenamefont {Hori},\ and\ \citenamefont
  {Sundholm}(2022)}]{nordlund2022large}%
  \BibitemOpen
  \bibfield  {author} {\bibinfo {author} {\bibfnamefont {K.}~\bibnamefont
  {Nordlund}}, \bibinfo {author} {\bibfnamefont {M.}~\bibnamefont {Hori}}, \
  and\ \bibinfo {author} {\bibfnamefont {D.}~\bibnamefont {Sundholm}},\
  }\bibfield  {title} {\enquote {\bibinfo {title} {Large nuclear scattering
  effects in antiproton transmission through polymer and metal-coated foils},}\
  }\href@noop {} {\bibfield  {journal} {\bibinfo  {journal} {Physical Review
  A}\ }\textbf {\bibinfo {volume} {106}},\ \bibinfo {pages} {012803} (\bibinfo
  {year} {2022})}\BibitemShut {NoStop}%
\bibitem [{\citenamefont {smarAct}(7 21)}]{piezo_drive}%
  \BibitemOpen
  \bibfield  {author} {\bibinfo {author} {\bibnamefont {smarAct}},\ }\href@noop
  {} {\emph {\bibinfo {title} {Note1, smarAct SLC-1730-CR -UHVP-NM-TI}}},\
  \bibinfo {organization} {smarAct} (\bibinfo {year} {Accessed:
  2022-07-21})\BibitemShut {NoStop}%
\bibitem [{\citenamefont {Kretzschmar}(2008)}]{kretzschmar2008calculating}%
  \BibitemOpen
  \bibfield  {author} {\bibinfo {author} {\bibfnamefont {M.}~\bibnamefont
  {Kretzschmar}},\ }\bibfield  {title} {\enquote {\bibinfo {title} {Calculating
  damping effects for the ion motion in a penning trap},}\ }\href@noop {}
  {\bibfield  {journal} {\bibinfo  {journal} {The European Physical Journal D}\
  }\textbf {\bibinfo {volume} {48}},\ \bibinfo {pages} {313--319} (\bibinfo
  {year} {2008})}\BibitemShut {NoStop}%
\bibitem [{\citenamefont {Gabrielse}\ \emph {et~al.}(1989)\citenamefont
  {Gabrielse}, \citenamefont {Fei}, \citenamefont {Orozco}, \citenamefont
  {Rolston}, \citenamefont {Tjoelker}, \citenamefont {Trainor}, \citenamefont
  {Haas}, \citenamefont {Kalinowsky},\ and\ \citenamefont
  {Kells}}]{gabrielse1989barkas}%
  \BibitemOpen
  \bibfield  {author} {\bibinfo {author} {\bibfnamefont {G.}~\bibnamefont
  {Gabrielse}}, \bibinfo {author} {\bibfnamefont {X.}~\bibnamefont {Fei}},
  \bibinfo {author} {\bibfnamefont {L.}~\bibnamefont {Orozco}}, \bibinfo
  {author} {\bibfnamefont {S.}~\bibnamefont {Rolston}}, \bibinfo {author}
  {\bibfnamefont {R.}~\bibnamefont {Tjoelker}}, \bibinfo {author}
  {\bibfnamefont {T.}~\bibnamefont {Trainor}}, \bibinfo {author} {\bibfnamefont
  {J.}~\bibnamefont {Haas}}, \bibinfo {author} {\bibfnamefont {H.}~\bibnamefont
  {Kalinowsky}}, \ and\ \bibinfo {author} {\bibfnamefont {W.}~\bibnamefont
  {Kells}},\ }\bibfield  {title} {\enquote {\bibinfo {title} {Barkas effect
  with use of antiprotons and protons},}\ }\href@noop {} {\bibfield  {journal}
  {\bibinfo  {journal} {Physical Review A}\ }\textbf {\bibinfo {volume} {40}},\
  \bibinfo {pages} {481} (\bibinfo {year} {1989})}\BibitemShut {NoStop}%
\end{thebibliography}%

\end{document}